\documentclass[]{article}
\usepackage{authblk}
\usepackage{graphicx}
\usepackage{amsfonts}
\usepackage[section]{placeins}

\title{Schwarzschild Black Hole in Anti-De Sitter Space} 
\author[]{M. Socolovsky$ ^* $}
\affil[]{Instituto de Ciencias, Universidad Nacional de General Sarmiento\newline \textit{Juan Mar\'\i a Guti\'errez 1150 (B1613GSX), Los Polvorines, Pcia. de Buenos Aires, Argentina}}
\date{}

\begin{document}
	\maketitle

\	
	
{\bf Abstract.} {\it We review several aspects of anti-De Sitter (AdS) spaces in different dimensions, and of four dimensional Schwarzschild anti-De Sitter (SAdS) black hole.}

\

\section{Introduction}

\

Anti-De Sitter spacetime [1] ($AdS_n$ in its different dimensions $n$) is a crucial ingredient in the formulation of the AdS/CFT conjecture [2]. Besides being in itself a solution of vacuum Einstein equations in the presence of a negative cosmological constant (maximal symmetric space with negative constant curvature and Lorentzian signature) [3], it is also an interesting laboratory for the study of black holes in a non asymptotically flat spacetime [4],[5],[6]. In particular the Schwarzschild case offers the possibility to discuss singularities, horizons and boundaries in a simple but non trivial way. An interesting aspect of their universal covering spacetimes $\widetilde{AdS_n}$ is that they consist of an infinite ``tower" in the time direction of their Euclidean constant negative curvature ``cousins" $HP^{n-1}$, the hyperbolic spaces. Timelike and lightlike geodesics have a curious behavior in these spacetimes [7],[8]. 

\

We divide our presentation basically in three parts. Starting with the hyperbolic (Lobachevsky) plane in its different versions, we then discuss in detail anti-De Sitter spacetime and its universal covering. As indicated in the Contents, we discuss symmetries and boundaries in all dimensions, and construct the Penrose diagram for $\widetilde{AdS_4}$. The largest part of the article is devoted to the four dimensional Schwarzschild anti-De Sitter black hole ($S\widetilde{AdS_4}$) which has two length scales: the mass $M$ of the black hole and the curvature radius $a$ of the embedding anti-De Sitter. In Schwarzschild coordinates we deduce the metric, give and explicit expression for the horizon $r_h=r_h(M,a)$, determine the surface gravity $\kappa$, and through the Rindler approximation [9],[10] near the horizon and the Unruh effect [11] (consequence of the Equivalence Principle), we find the Hawking temperature [12]. A careful analysis of the tortoise or Regge-Wheeler radial coordinate $r^*$ [13] allows to define without any ambiguity the Kruskal-Szekeres coordinates [14],[15],[16] and then construct the Penrose diagram. We end the review with a detailed derivation of the thermodynamic energy and the area law for the entropy of $S\widetilde{AdS_4}$, and a brief description of the Hawking-Page phase transition.

\

We use natural units $\hbar=c=G_N=k_B=1$.

\

\section{The hyperbolic plane (Lobachevsky plane)}

Since the $\widetilde{AdS}_n$, $n$=2,3,4,...spacetimes can be seen as a continuous stack in the time direction of their Euclidean counterparts in one less dimension, the hyperbolic spaces $H^{n-1}$, we start this review with a systematic construction of the geometry of the hyperbolic plane $H^2$ related to $\widetilde{AdS}_3$ in different coordinate systems (five in total), coordinate systems which are later used in the construction of the $AdS$ spacetimes, adding of course the time direction and passing to a Lorentzian signature of the metric. The generalization to higher dimensions is quite trivial and, though it is exhibited in eqs. (2.35) and (3.44), it is not derived in detail.

\

2.1 Take the pseudoeuclidean space $\mathbb{E}^{(2,1)}$ with coordinates $X,Y,Z\in (-\infty,+\infty)$ and metric $$dl^2=dX^2+dY^2-dZ^2.\eqno{(2.1)}$$ $[X]=[Y]=[Z]=L^1.$

\

\begin{figure}[h!]\centering
	\includegraphics[width=.7\linewidth]{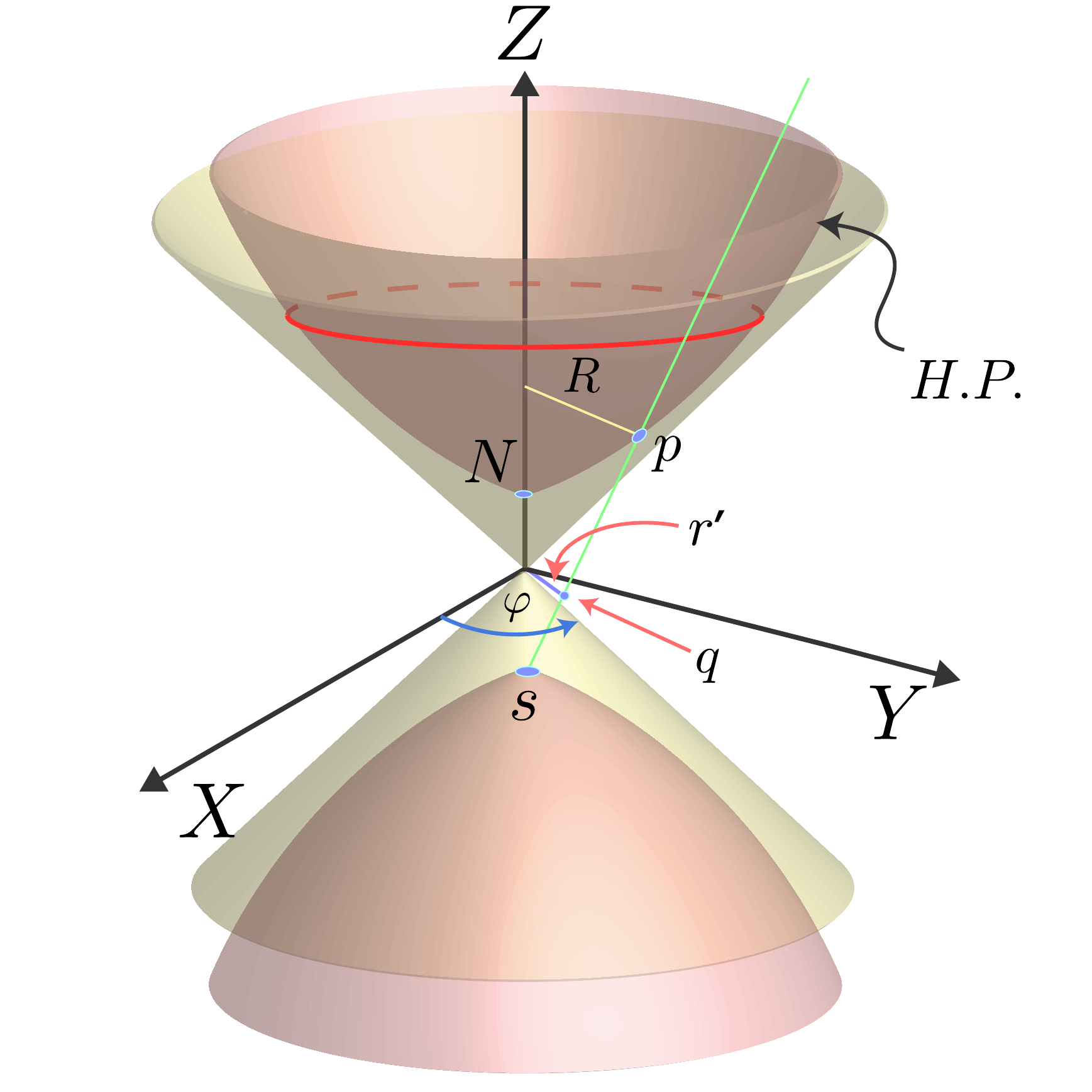}
	\caption{Hyperbolic plane.}
	\label{f1}
\end{figure}

2.2.  The {\it hyperbolic plane} ($H.P.$) is defined by the upper (or lower) half of the 2-sheet hyperboloid $$X^2+Y^2-Z^2=-a^2, \eqno{(2.2)}$$ $a>0$, $[a]=L^1$. (See Fig. \ref{f1}.)

2.3. Parametrize $X$, $Y$ and $Z$ as follows: $$X:=a \ Sh\rho \  cos\varphi, \ Y:=a \ Sh\rho \ sin\varphi, \ Z:=a \ Ch\rho\eqno{(2.3)}$$ with $$\rho\in [0,+\infty), \ \varphi\in [0,2\pi), \ [\rho]=[\varphi]=L^0.\eqno{(2.3a)}$$ Spatial infinity is at $\rho=+\infty$ while spatial origin $o$ is at $\rho=0$: $(X_o,Y_o,Z_o)=(0,0,a)\equiv N$. 

\

2.4. $\rho$ and $\varphi$ can be taken as coordinate functions for $H.P.$ since (2.3a) satisfies (2.2): $$l.h.s.{(2.2)}=a^2 \ Sh^2\rho \ cos^2\varphi +a^2 \ Sh^2\rho \ sin^2\varphi-a^2 \ Ch^2\rho$$ $$=a^2(Sh^2\rho -Ch^2\rho)=-a^2.$$

\

2.5. Embedding of the 2-sheet hyperboloid (2.2) in $\mathbb{E}^{(2,1)}$: from (2.2), $$Z=\pm\sqrt{X^2+Y^2+a^2}=Z(X,Y,a).\eqno{(2.4)}$$ So, $Z(0,0,a)=\pm a$, $Z(0,Y,a)=\pm\sqrt{Y^2+a^2}\to \pm Y$ as $Y\to\pm\infty$, $Z(X,0,a)=\pm\sqrt{X^2+a^2}\to \pm X$ as $X\to\pm\infty$.

\

We called $(0,0,a)=N=o$ (origin): north pole of the hyperboloid, origin of $H.P.$ Also, call $(0,0,-a)\equiv S$ (south pole of the hyperboloid). 

\

2.6. The metric (2.1) in $\mathbb{E}^{(2,1)}$ induces a metric in $H.P.$: $$dX=a \ Ch\rho \ \cos\varphi \ d\rho-a \ Sh\rho \ sin\varphi \ d\varphi,$$ $$dY=
 a \ Ch\rho \ \sin\varphi \ d\rho+a \ Sh\rho \ cos\varphi \ d\varphi$$ and $$dZ=a \ Sh\rho \ d\rho$$ imply $$dl^2_{H.P.}=(dX^2+dY^2-dZ^2)_{H.P.}=a^2(d\rho^2+Sh^2\rho \ d\varphi^2).\eqno{(2.5)}$$ I.e. the H.P. is a space (not a spacetime) with metric $${g_{ij}}_{H.P.}=\pmatrix{g_{\rho\rho} & g_{\rho\varphi}\cr g_{\varphi\rho}& g_{\varphi\varphi}\cr}=a^2\pmatrix{1 & 0\cr 0 & Sh^2\rho\cr}.\eqno{(2.6)}$$ At $\rho=0$, $det({g_{ij}}_{H.P.})=0$ and then it does not exist ${g^{ij}}_{H.P.}(0,\varphi)$. We emphasize that $\rho =0$ is only a coordinate singularity.
 
 \
 
 {\it Note}: The metric is also valid for the lower (or upper) hyperboloid.
 
 \
 
 2.7. {\it Poincar\'e projection ({\it Poincar\'e disk}) of $H.P.$: $H.P.|_{Poinc.}$}

\

A straight line from $S$ to a point $p$ of $H.P.$ crosses the $X-Y$ plane at a point $q$. Let $R$ and $r^\prime$ be the respective distances to the $Z$-axis from $p$ and $q$. From (2.3), $R=\sqrt{X^2+Y^2}=a \ Sh\rho$, and from $${{r^\prime}\over{a}}={{R}\over{Z+a}}\eqno{(2.7)}$$ one obtains $$r^\prime=a{{Sh\rho}\over{1+Ch\rho}}\eqno{(2.8)}$$ i.e. $r^\prime=r^\prime(\rho)$, with $r^\prime(0)=0$, $r^\prime(\rho)<a$ since ${{Sh\rho}\over{1+Ch\rho}}<1$ and $$r^\prime(\rho)\to a_- \ \ as \ \ \rho\to +\infty. \eqno{(2.9)}$$

\

$dr^\prime={{dr^\prime}\over{d\rho}}d\rho$ and ${{1}\over{a}}{{dr^\prime}\over{d\rho}}={{1}\over{1+Ch\rho}}$; then $$dr^\prime={{a}\over{1+Ch\rho}}d\rho, \ \ {dr^\prime}^2={{a^2}\over{(1+Ch\rho)^2}}d\rho^2.\eqno{(2.10)}$$

\

So, for $\rho>>1$, $$dr^\prime\cong 2ae^{-\rho}d\rho\to 0 \ \ as \ \ \rho\to +\infty.\eqno{(2.11)}$$ The metric for $H.P.|_{Poinc.}$ (in the $X,Y$ plane) is $$dl^2_{Poinc.}={dr^\prime}^2+{r^\prime}^2d\varphi^2={{1}\over{(1+Ch\rho)^2}}a^2(d\rho^2+Sh^2\rho \ d\varphi^2)={{1}\over{(1+Ch\rho)^2}}dl^2_{H.P.},\eqno{(2.12)}$$ i.e. the Poincar\'e disk and the hyperbolic plane are conformally equivalent: $$H.P.|_{Poinc.}\buildrel \ {conf.}\over\cong H.P.\eqno{(2.13)}$$

\

2.8. {\it First change of coordinates} (for $H.P.$, eq.(2.5))

\

$$(\rho,\varphi)\to (\chi,\varphi), \ \ Ch\rho:={{1}\over{cos\chi}}, \ \ \chi\in [0,\pi/2).\eqno{(2.14)}$$ From $Ch^2\rho-Sh^2\rho=1$ one obtains $Sh\rho=tg\chi$, and ${{dCh\rho}\over{d\rho}}=Sh\rho$ implies $d\rho={{d\chi}\over{cos\chi}}$; then, $$dl^2_{H.P.}={{a^2}\over{cos^2\chi}}(d\chi^2+sin^2\chi d\varphi^2).\eqno{(2.15)}$$ (This expression will be used to show that $AdS_3$ is a stack of $H.P.$'s.)

\

2.9. {\it Second change of coordinates}

\

$$(\chi,\varphi)\to (r,\varphi), \ \ r:=a \ tg\chi, \ \ r\in[0,+\infty), \ \ [r]=L^1.\eqno{(2.16)}$$ From $d\chi={{dr}\over{a(1+r^2/a^2)}}$ one obtains $$dl^2_{H.P.}={{dr^2}\over{1+r^2/a^2}}+r^2d\varphi^2.\eqno{(2.17)}$$

\

2.10. {\it Third coordinate change: Poincar\'e coordinates} 

\

Define the coordinates $$\iota:= a \ ln({{Y+Z}\over{a}}), \ \tau:={{X}\over{Y+Z}}; \ \iota,\tau\in (-\infty,+\infty), \ [\iota]=L^1, \ [\tau]=L^0.\eqno{(2.18)}$$ From (2.2) it can be seen that $Y>-Z$ corresponds to the upper hyperboloid. In these coordinates, $$dl^2_{H.P.}=a^2 \ e^{2\iota /a}d\tau^2+d\iota^2.\eqno{(2.19)}$$ Let $$s:=a \ e^{\iota /a}; \ s\in (0,+\infty), \ (s(+\infty)=+\infty, \ s(0)=a, \ s(-\infty)=0), \ [s]=L^1.\eqno{(2.20)}$$ The metric becomes $$dl^2_{H.P.}=s^2d\tau^2+a^2{{ds^2}\over{s^2}}.\eqno{(2.21)}$$

\

2.11. {\it Fourth coordinate change: Poincar\'e half plane (P.H.P.)}

\

With $s:={{a}\over{z}}$, $z\in(0,+\infty)$, $[z]=L^0$, $$dl^2_{H.P.}={{a^2}\over{z^2}}(d\tau^2+dz^2).\eqno{(2.22)}$$ Finally, defining $$x:=a\tau\in(-\infty,+\infty), \ y:=az\in(0,+\infty); [x]=[y]=L^1, \eqno{(2.23a)}$$ the metric of the $H.P.$ or $P.H.P.$ is $$dl^2_{H.P.}={{a^2}\over{y^2}}(dx^2+dy^2).\eqno{(2.23b)}$$ (See Fig. 2.) So, the $P.H.P.$ is conformally equivalent to the upper half plane in $\mathbb{E}^2$. 

\

The metric tensor is $${g_{ij}}_{P.H.P.}=a^2\pmatrix{g_{xx} & g_{xy} \cr g_{yx} & g_{yy}\cr}={{a^2}\over{y^2}}\pmatrix{1 & 0\cr 0 & 1\cr},\eqno{(2.24a)}$$ with inverse $${g^{ij}}_{P.H.P.}={{y^2}\over{a^2}}\pmatrix{1 & 0\cr 0 & 1\cr}.\eqno{(2.24b)}$$ We want to stress here that the $P.H.P.$ and the $H.P.$ are both topologically and geometrically equivalent. 

\begin{figure}[h!]\centering
	\includegraphics[width=\linewidth]{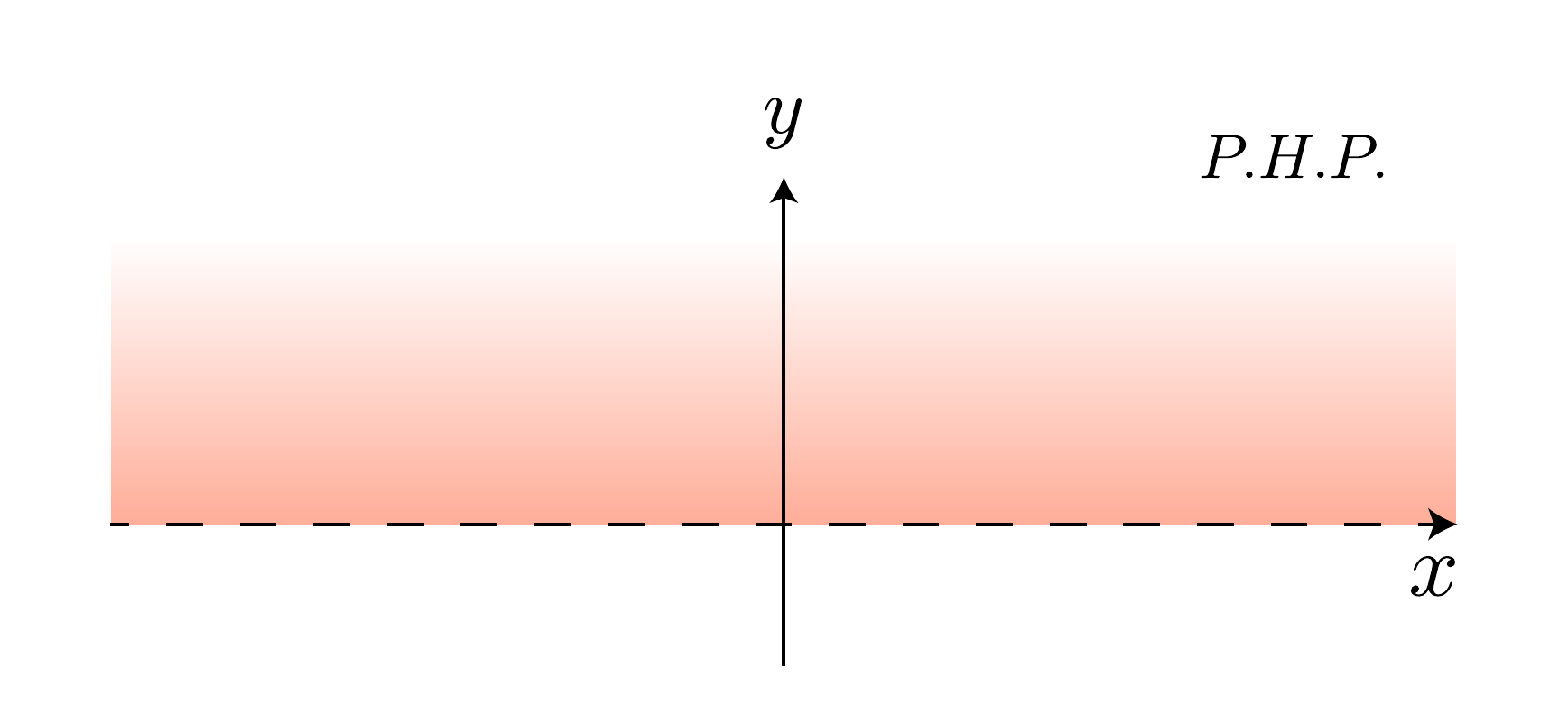}
	\caption{Poincar\'e half plane.}
	\label{f2}
\end{figure}

2.12. {\it Scalar curvature}

\

To calculate it we use the $P.H.P.$ coordinates $x,y$. 

\

i) The Christoffel symbols are given by $$\Gamma^\mu_{\nu\rho}={{1}\over{2}}g^{\mu\lambda}(\partial_\nu g_{\rho\lambda}+\partial_\rho g_{\nu\lambda}-\partial_\lambda g_{\nu\rho})\eqno{(2.25)}$$ with $\mu,\nu,...\in \{x,y\}$, $[\Gamma^\mu_{\nu\rho}]=L^{-1}$. A straightforward calculation gives $$\Gamma^x_{xy}=\Gamma^x_{yx}=\Gamma^y_{yy}=-\Gamma^y_{xx}=-{{1}\over{y}},\eqno{(2.25a)}$$ the other symbols being zero. 

\

ii) For the Riemann curvature tensor one has $$R_{\mu\nu\rho\sigma}=(g_{\mu\rho}g_{\nu\sigma}-g_{\mu\sigma}g_{\nu\rho}){{R_{xyxy}}\over{g}}, \ g=det(g_{\mu\nu})={{a^4}\over{y^4}}.$$ Then for the Ricci tensor one obtains $$R_{\nu\sigma}=R^\mu_{\nu\mu\sigma}={{R_{xyxy}}\over{g}}g_{\nu\sigma}$$ with scalar curvature $$R=R^\nu_\nu={{2y^4}\over{a^4}}R_{xyxy}.\eqno{(2.26)}$$ 

iii) From the definition $$R^\rho_{\sigma\mu\nu}=\partial_\mu\Gamma^\rho_{\sigma\nu}-\partial_\nu\Gamma^\rho_{\sigma\mu}+\Gamma^\lambda_{\sigma\nu}\Gamma^\rho_{\lambda\mu}-\Gamma^\lambda_{\sigma\mu}\Gamma^\rho_{\lambda\nu},$$ $R^x_{yxy}=-{{1}\over{y^2}}$ and so $R_{xyxy}=g_{xx}R^x_{yxy}=-{{a^2}\over{y^4}}$. Then $$R=-{{2}\over{a^2}}. \eqno{(2.27)}$$ $R<0$, $[R]=L^{-2}$, and $R\to 0_-$ as $a\to +\infty$ ($R\to -\infty$ as $a\to 0_+$).

\

2.13. {\it Vertical distance between two points: $(x,y_1)$ and $(x,y_2)$, $y_2>y_1$}

\

$dx=0$ implies $dl^2_{H.P.}={{a^2}\over{y^2}}dy^2$ and so $$\Delta l=a\int^{y_2}_{y_1}{{dy}\over{y}}=a \ ln({{y_2}\over{y_1}}). \eqno{(2.28)}$$ So, $\Delta l\to +\infty$ as $y_1\to 0_+$ i.e. $y=0$ is infinitely far away: it is a boundary at spatial infinity, like $z=0$ in $AdS$ (see (3.31)).

\

2.14. {\it Horizontal distance between two points: $(x_1,y)$ and $(x_2,y)$, $x_2>x_1$}

\

$dy=0$ implies $dl^2_{H.P.}={{a^2}\over{y^2}}dx^2$ and so $$\Delta l={{a}\over{y}}(x_2-x_1).\eqno{(2.29)}$$ So, for fixed $x_2-x_1$, $\Delta l\to 0$ as $y\to +\infty$ and $\Delta l\to +\infty$ as $y\to 0_+$. 

\

2.15. {\it 3-dimensional hyperbolic space $H^3$}

\

Take the pseudo-Euclidean space $\mathbb{E}^{(3,1)}$ with metric $$dl^2_{\mathbb{E}^{(3,1)}}=dx^2+dy^2+dw^2-dz^2.\eqno{(2.30)}$$ $H^3$ is the 3-dimensional hyperboloid (upper: $z>0$ or lower: $z<0$), subspace of $\mathbb{E}^{(3,1)}$, defined by $$x^2+y^2+w^2-z^2=-a^2,$$ $$x,y,w,z\in(-\infty,+\infty), \ [x]=[y]=[w]=[z]=[a]=L^1, \ a>0.\eqno{(2.31)}$$ In terms of the parameters $$\rho\in[0,+\infty), \ \theta\in[0,\pi], \ \varphi\in[0,2\pi), \ [\rho]=[\theta]=[\varphi]=L^0, \eqno{(2.32)}$$ $$x=a \ Sh\rho \ cos\theta,$$ $$y=a \ Sh\rho \ sin\theta \ cos\varphi,$$ $$w=a \ Sh\rho \ sin\theta \ sin\varphi,$$ $$z=a \ Ch\rho\eqno{(2.33)}$$ satisfy (2.31) with metric $$dl^2_{H^3}=a^2(d\rho^2+Sh^2\rho \ d\Omega_2^2),\eqno{(2.34)}$$ where $d\Omega_2^2=d\theta^2+sin^2\theta \ d\Omega_1^2$, $d\Omega_1^2=d\varphi^2$. $\rho$, $\theta$ and $\varphi$ are declared {\it global coordinate functions} on $H^3$. With the change of coordinates (2.14), the metric becomes $$dl^2_{H^3}={{a^2}\over{cos^2\chi}}(d\chi^2+sin^2\chi \ d\Omega_2^2).\eqno{(2.35)}$$ 

\

\section{Four dimensional anti-De Sitter spacetime ($AdS_4$) and its universal covering $\widetilde{AdS_4}$}

\

3.1. Consider the pseudo-Euclidean space $\mathbb{E}^{(2,3)}$ with global coordinates $x^\mu=t,v,x,y,z$, 

\

$x^\mu\in (-\infty,+\infty)$, $[x^\mu]=L^1$ and metric $$ds^2=dt^2+dv^2-dx^2-dy^2-dz^2. \eqno{(3.1)}$$

\

3.2. The {\it 4-dimensional anti-De Sitter spacetime} $AdS_4\subset\mathbb{E}^{(2,3)}$ is defined by $$t^2+v^2-x^2-y^2-z^2=a^2, \ a=const.<0, \ [a]=L^1.\eqno{(3.2)}$$ 

\

({\it Visualization}: In 2 dimensions, $AdS_2\subset\mathbb{E}^{(2,1)}$ with metric $ds^2=dt^2+dv^2-dx^2$ is given by the 1-sheet hyperboloid $x=\pm\sqrt{t^2+v^2-a^2}$ around the $x$-axis. See Fig. 3.)

\begin{figure}[h!]\centering
	\includegraphics[width=.7\linewidth]{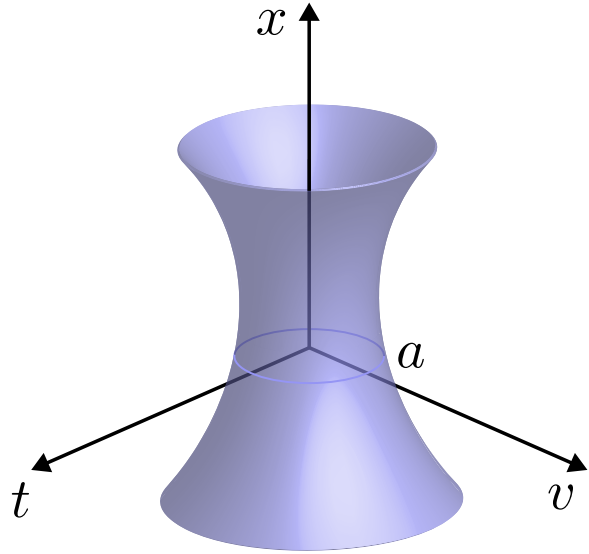}
	\caption{2-dim. anti-De Sitter space}
	\label{f3}
\end{figure}

3.3. One defines the four parameters $y^\mu=t^\prime,\rho,\theta,\varphi$, $[y^\mu]=L^0$, $t^\prime\in[0,2\pi]$, $\rho\in[0,+\infty)$, $\theta\in[0,\pi]$, $\varphi\in[0,2\pi)$ on which the $x^\mu$'s depend and obey (3.2) through $$x=a \ Sh\rho \ cos\theta, \ y=a \ Sh\rho \ sin\theta \ cos\varphi, \ z=a \ Sh\rho \ sin\theta \ sin\varphi,$$ $$t=a \ Ch\rho \ sin \ t^\prime, \ v=a \ Ch\rho \ cos \ t^\prime.\eqno{(3.3)}$$

\

3.4. Replacing (3.3) in (3.1) one obtains $$ds^2_{AdS_4}=a^2(Ch^2\rho \ d{t^\prime}^2-d\rho^2-Sh^2\rho \ d\Omega^2_2). \eqno{(3.4)}$$ At this step one declares the set $\{\rho,t^\prime,\theta,\varphi\}$ as {\it global coordinate functions} on $AdS_4$, with $\rho$: radial coordinate, $t^\prime$: time coordinate, and $\theta$ and $\varphi$ angular coordinates. 

\

Notice that $g_{t^\prime t^\prime}=<\partial_{t^\prime},\partial_{t^\prime}>=a^2 \ Ch^2\rho >0$ for all $\rho\geq 0$ i.e. $\partial_{t^\prime}$ is a timelike Killing vector field for all $t^\prime$ which is periodic; then one should have {\it closed timelike curves}. To avoid them one makes the {\it extension} $$t^\prime\in [0,2\pi]\to t^\prime\in(-\infty,+\infty),\eqno{(3.4a)}$$ i.e. one unwraps the circle $S^1$, passing to the {\it universal covering space} of $AdS_4$, $\widetilde{AdS_4}$.

\

3.5. {\it Symmetry groups}

\

$Symm(\widetilde{AdS_4})=SO(2,3)$ (analogously as $Symm(S^2)=SO(3)$). Clearly, $\widetilde{AdS_4}$ is not translation invariant i.e. $T_{(2,3)}$ is not a symmetry. In general, $Symm(\widetilde{AdS_d})=SO(2,d-1)$; e.g. $Symm(\widetilde{AdS_5})=SO(2,4)$. Notice that $dim_\mathbb{R}(SO(2,4))=15=dim_\mathbb{R}(Conf(Mink^4))$. ($Conf$ denotes the {\it conformal group} and $Mink$ is Minkowski spacetime.) In general, $dim_\mathbb{R}(Symm(\widetilde{AdS_d}))=dim_\mathbb{R}(SO(2,d-1))={{(d+1)d}\over{2}}=dim_\mathbb{R}(Conf(Mink^{d-1}))$ for $d\geq 4$. (See subsection 3.7.)

\

3.6. {\it Change of coordinates: 3rd. coordinate system; conformal Penrose-Carter diagram}

\

$$(t^\prime,\rho,\theta,\varphi)\to (t^\prime,\chi,\theta,\varphi), \ Ch\rho:={{1}\over{cos\chi}}, \ \chi\in[0,\pi/2).\eqno{(3.5)}$$ $\chi=0$ corresponds to $\rho=0$: spatial origin, while $\chi\to\pi/2_-$ corresponds to $\rho\to +\infty$: spatial infinity. So, the change $\rho\to\chi$ ``brings" spatial infinity to finite ``distance". The metric becomes $$ds^2_{\widetilde{AdS_4}}={{a^2}\over{cos^2\chi}}({dt^\prime}^2-d\chi^2-sin^2\chi \ d\Omega^2_2).\eqno{(3.6)}$$ For a 3-sphere of unit radius one has the metric $$ds^2_{S^3}=d\chi^2+sin^2\chi \ d\Omega^2_2, \eqno{(3.7)}$$ with $\chi\in[0,\pi]$, $\chi=0$ and $\chi=\pi$ respectively being the south and north poles of $S^3$. If $\chi$ in (3.6) should extend to $[0,\pi]$ then the round parenthesis would correspond to the Einstein static universe with topology $\mathbb{R}\times S^3$. However, since for $\widetilde{AdS_4}$, $\chi\in[0,\pi/2]$, it turns out that $\widetilde{AdS_4}$ {\it is conformally equivalent (with conformal factor $\omega(\chi)={{a}\over{cos\chi}}$) to half of Einstein static universe}. ``Half" corresponds to half hemisphere of $S^3$without boundary which, topologically, is $\mathbb{R}^3$. Then $$\widetilde{AdS_4}\buildrel \ {top.}\over\cong \mathbb{R}\times\mathbb{R}^3\buildrel \ {top.}\over\cong \mathbb{R}^4.\eqno{(3.8)}$$ The {\it conformal Penrose-Carter diagram} is given in Fig. 4. $p$ is a point at spatial origin ($\chi=0$); $q$ is a ``point" at spatial infinity ($\chi=\pi/2$): in fact $q$ is a 2-sphere at infinity of radius $sin(\pi/2)=1$ with $\{\chi=\pi/2\} \buildrel \ {top.} \over\cong\mathbb{R}\times S^2$, timelike hypersurface. Notice that $\{\chi=\pi/2\}\not\subset\widetilde{AdS_4}$. The radial ($d\theta=d\varphi=0$) light signals are given by $ds^2_{\widetilde{AdS_4}}={dt^\prime}^2-d\chi^2=0$ i.e. by the straight lines at $\pm 45^o$, ${{dt\prime}\over{d\chi}}=\pm 1$. $\times a$ denotes a radial light cone at point $a$. 

\begin{figure}[h!]\centering
	\includegraphics[width=.7\linewidth]{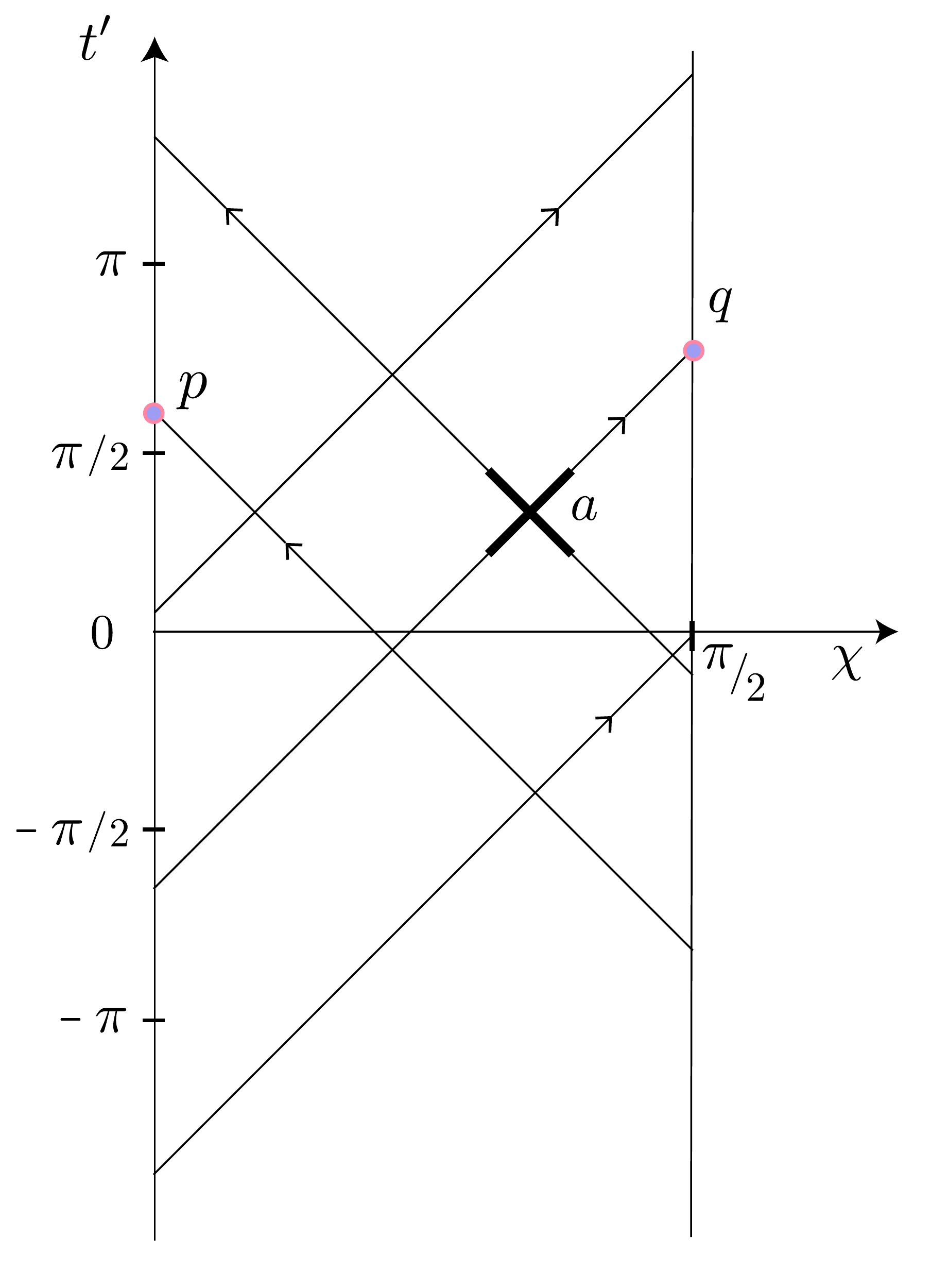}
	\caption{Conformal Penrose-Carter diagram of $ \widetilde{AdS_4} $}
\end{figure}

3.7. {\it $\partial(\widetilde{AdS_4})$: boundary of the universal covering space of $AdS_4$}

\

For large $\rho$ (towards spatial infinity), $Ch\rho$, $Sh\rho$ $\to {{e^\rho}\over{2}}$ and so, from (3.4), $$ds^2_{\widetilde{AdS_4}}\buildrel \ {\rho>>1}\over\longrightarrow {{e^{2\rho}a^2}\over{4}}({dt^\prime}^2-d\Omega^2_2).$$ I.e. $\partial(\widetilde{AdS_4})$ is geometrically conformal to $\mathbb{R}\times S^2$ and therefore $$\partial(\widetilde{AdS_4})\buildrel \ {top.}\over\cong\mathbb{R}\times S^2,\eqno{(3.9)}$$ while $$\partial(AdS_4)\buildrel \ {top.}\over\cong S^1\times S^2.\eqno{(3.9a)}$$ In general, $$\partial(\widetilde{AdS_d})\buildrel \ {top.}\over\cong\mathbb{R}\times S^{d-2},\eqno{(3.10)}$$ while  $$\partial(AdS_d)\buildrel \ {top.}\over\cong S^1\times S^{d-2}.\eqno{(3.10a)}$$ As another example, $$\partial(\widetilde{AdS_5})\buildrel \ {top.}\over\cong\mathbb{R}\times S^3,\eqno{(3.11)}$$ while $$\partial(AdS_5)\buildrel \ {top.}\over\cong S^1\times S^3.\eqno{(3.11a)}$$ But $S^1\times S^3 \buildrel \ {top.}\over\cong U(1)\times SU(2)$, the {\it electroweak group}, so $$\partial(AdS_5)\buildrel \ {top.}\over\cong U(1)\times SU(2). \eqno{(3.12)}$$

\

On the other hand, $S^{d-2}=\mathbb{R}^{d-2}\cup\{\infty\}={(\mathbb{R}^{d-2})}^c$: one point compactification of $\mathbb{R}^{d-2}$. But $\mathbb{R}^{d-2}\buildrel \ {top.}\over\cong s.p.(Mink^{d-1})$ ($s.p.$: spatial part); then $$S^{d-2}\buildrel \ {top.}\over\cong {(s.p.(Mink^{d-1}))}^c,\eqno{(3.13)}$$ and therefore $$\partial(\widetilde{AdS_d})\buildrel \ {top.}\over\cong\mathbb{R}\times {(s.p.(Mink^{d-1}))}^c,\eqno{(3.14)}$$ while $$\partial(AdS_d)\buildrel \ {top.}\over\cong S^1\times {(s.p.(Mink^{d-1}))}^c.\eqno{(3.14a)}$$ In particular, $$\partial(\widetilde{AdS_4})\buildrel \ {top.}\over\cong\mathbb{R}\times {(s.p.(Mink^3))}^c,\eqno{(3.15)}$$ while $$\partial(AdS_4)\buildrel \ {top.}\over\cong S^1\times {(s.p.(Mink^3))}^c.\eqno{(3.15a)}$$

\

3.8. {\it 4th coordinate system: ``spherical or static coordinates"}

\

If in (3.6) we define $$r:=a \ tg\chi; \ r\in [0,+\infty), \ \chi=0\Rightarrow r=0, \ \chi\to\pi/2_-\Rightarrow r\to+\infty,$$ $$t:=at^\prime; \ t\in(-\infty,+\infty), [r]=[t]=L^1,\eqno{(3.16)}$$ one obtains $$ds^2_{\widetilde{AdS_4}}=(1+{{r^2}\over{a^2}})dt^2-{{dr^2}\over{1+r^2/a^2}}-r^2d\Omega^2_2.\eqno{(3.17)}$$ $t$, $r$, $\theta$ and $\varphi$ are global coordinates on $\widetilde{AdS_4}$. The same expression gives $ds^2_{AdS_4}$ but with $t\in [-\pi a,+\pi a]$.

\

Using $cos\chi=1/\sqrt{1+tg^2\chi}$ one easily finds the relation between $r$ and the radial coordinate $\rho$ in (3.4): $$r=a \ Sh\rho. \eqno{(3.18)}$$ So, $r\to\infty$ is spatial infinity. 

\

3.9. A straightforward calculation leads to the {\it scalar curvature} $R$ of $\widetilde{AdS_4}$ (or $AdS_4$): $$R=-{{6}\over{a^2}}.\eqno{(3.19)}$$ The {\it cosmological constant} is defined by $$\Lambda:={{1}\over{2}}R=-{{3}\over{a^2}}<0.\eqno{(3.20)}$$ In terms of $\Lambda$, $$ds^2_{\widetilde{AdS_4}}=(1-\Lambda r^2/3)dt^2-{{dr^2}\over{1-\Lambda r^2/3}}-r^2d\Omega^2_2.\eqno{(3.21)}$$ So, $\widetilde{AdS_4}$ is the universal covering spacetime of the maximally symmetric 4-dimensional solution to the Einstein equations with negative cosmological constant. (This fact can be seen from eq. (4.11) in Section 4, with $M=0$, where $M$ is the Schwarzschild black hole mass parameter.) $\Lambda<0$ leads to an {\it attractive} gravitational force since if the $g_{tt}$ component of the metric is written as $1+2\Phi$, to the ``Newtonian" potential $\Phi={{r^2}\over{2a^2}}$ corresponds a ``force" $-{{d\Phi}\over{dr}}=-{{r}\over{a^2}}={{\Lambda}\over{3}}r$. So, spatial infinity behaves as an infinite potential wall. 

\

3.10. {\it 5th coordinate system}

\

If in (3.2) we call $$t=x_0, \ x=x_1, \ y=x_2, \ z=x_3, \ v=x_4,\eqno{(3.22)}$$ the equation defining $\widetilde{AdS_4}\subset\mathbb{E}^{(2,3)}$ is $$(x_0^2+x_4^2)-(x_1^2+x_2^2+x_3^2)=a^2.\eqno{(3.2a)}$$ In terms of the parameters $$Y:=a \ln({{x_4+x_3}\over{a}}), \ \tau:={{x_0}\over{x_4+x_3}}, \ y_1:={{x_1}\over{x_4+x_3}}, \ y_2:={{x_2}\over{x_4+x_3}},$$ $$Y, \ \tau, \ y_i \ (i=1,2) \ \in(-\infty,+\infty), \ [Y]=L^1, \ [\tau]=[y_i]=L^0, \eqno{(3.23)}$$ the coordinate functions $x_\mu$ are given by $$x_0=x_0(\tau,Y,y_i)=ae^{Y/a}\tau,$$ $$x_i=x_i(\tau,Y,y_i)=ae^{Y/a}y_i, \ i=1,2,$$ $$x_3=x_3(\tau,Y,y_i)=a \ Sh(Y/a)-{{a}\over{2}}e^{Y/a}(y_1^2+y_2^2-\tau^2),$$ $$x_4=x_4(\tau,Y,y_i)=a \ Ch(Y/a)+{{a}\over{2}}e^{Y/a}(y_1^2+y_2^2-\tau^2),\eqno{(3.24)}$$ and they obey (3.2a). Then, $\{\tau,Y,y_i\}$ are declared coordinate functions on $\widetilde{AdS_4}/2$ with metric  $$ds^2_{\widetilde{AdS_4}/2}=a^2e^{{{2Y}\over{a}}}(d\tau^2-dy_1^2-dy_2^2)-dY^2.\eqno{(3.25)}$$ $\widetilde{AdS_4}/2$ denotes the half spacetime, since $Y\in\mathbb{R}$ if and only if $$x_4+x_3>0. \eqno{(3.26)}$$ 

\

3.11. {\it 6th coordinate system: Poincar\'e coordinates}

\

3.11.a. {\it Poincar\'e coordinates}

\

Let $$r=r(Y):=ae^{Y/a}, \ r>0, \ [r]=L^1.\eqno{(3.27)}$$ Then: $r(0)=a \ (Y=0\Leftrightarrow x_4+x_3=a), \ r\to 0_+: \ {\it cosmological \ horizon} \ as \ Y\to -\infty \Leftrightarrow (x_4+x_3\to 0_+), \ r\to +\infty: \ {\it spatial \ infinity} \ as \ Y\to+\infty\Leftrightarrow x_4+x_3\to +\infty.$ The metric becomes $$ds^2_{\widetilde{AdS_4}/2}=r^2(d\tau^2-dy_1^2-dy_2^2)-a^2{{dr^2}\over{r^2}}.\eqno{(3.28)}$$ ({\it Note}: (3.25) and (3.28) are respectively analogous to (2.19) and (2.21) corresponding to the $H.P.$. We have here $d=4$ (then the additional coordinates $y_1$ and $y_2$) and a Lorentzian signature for time $\tau$.)

\

$ds^2_{\widetilde{AdS_4}/2}$ is invariant under the {\it scale transformation} $$\tau\to \lambda\tau, \ y_i\to\lambda y_i, \ r\to r/\lambda, \ \lambda >0.\eqno{(3.29)}$$ Finally, with $$r={{a}\over{z}}, \ z>0,\eqno{(3.30)}$$ the metric becomes $$ds^2_{\widetilde{AdS_4}/2}={{a^2}\over{z^2}}(d\tau^2-dy_1^2-dy_2^2-dz^2),$$ $$\tau,y_i\in (-\infty,+\infty), \ z\in (0,+\infty),$$ $$[\tau],=[y_i]=[z]=L^0, \ [a]=L^1. \eqno{(3.31)}$$ The {\it scale invariance} of the metric is like (3.29) for $\tau$ and $y_i$, with $z\to\lambda z$. (The metric (3.31) is analogous to the metric (2.22) for the P.H.P.)

\

(3.31) says us that $$\widetilde{AdS_4}/2 \buildrel \ {conf.}\over\cong Mink^4/2\buildrel \ {top.}\over\cong Mink^4,\eqno{(3.32)}$$ with conformal factor $\Omega(z)={{a}\over{z}}$.

\

3.11.b. {\it Generalization to arbitrary dimensions}

\

The generalization to the $d$-dimensional anti-De Sitter spacetime is straightforward with obvious definition of the additional coordinates. With $$x_d+x_{d-1}>0, \eqno{(3.33)}$$ the metric is $$ds^2_{\widetilde{AdS_d}/2}=a^2e^{2Y/a}(d\tau^2-\sum_{i=1}^{d-2}dy_i^2)-dY^2$$ $$=r^2(d\tau^2-\sum_{i=1}^{d-2}dy_i^2)-a^2{{dr^2}\over{r^2}}$$ $$={{a^2}\over{z^2}}(d\tau^2-\sum_{i=1}^{d-2}dy_i^2-dz^2).\eqno{(3.34)}$$ The metric can be expressed in coordinates with the more familiar dimension of length: $$ds^2_{\widetilde{AdS_d}/2}={{a^2}\over{Z^2}}(dT^2-\sum_{i=1}^{d-2}dY_i^2-dZ^2),\eqno{(3.35)}$$ with $$T=a\tau, \ Y_i=ay_i, \ Z=az, \ T,Y_i\in(-\infty,+\infty), \ Z\in(0,+\infty), \ [T]=[Y_i]=[Z]=L^1.$$ The generalization of (3.6) and (3.17) are, respectively, $$ds^2_{\widetilde{AdS_d}}={{a^2}\over{cos^2\chi}}({dt^\prime}^2-d\chi^2-sin^2\chi \ d\Omega^2_{d-2}).\eqno{(3.6a)}$$ and $$ds^2_{\widetilde{AdS_d}}=(1+{{r^2}\over{a^2}})dT^2-{{dr^2}\over{1+r^2/a^2}}-r^2d\Omega^2_{d-2}.\eqno{(3.17a)}$$ As for the 4-dimensional case, $$\widetilde{AdS_d}/2\buildrel \ {conf.}\over\cong Mink^d/2\buildrel \ {top.}\over\cong Mink^d.\eqno{(3.36)}$$

\

The ``other half'' of $\widetilde{AdS_d}$ corresponds to $$x_d+x_{d-1}=-ae^{Y/a}<0\eqno{(3.37)}$$ with $$x_d+x_{d-1}\to -\infty \ and \ r=ae^{Y/a}\to +\infty \ as \ Y\to +\infty.\eqno{(3.38)}$$ Fig. 5 illustrates the two patches of $\widetilde{AdS_d}$ with respectively unprimed and primed coordinates for the ``right'' and ``left'' patches. The expression for the metric is the same for both patches, except for the sign of $x_d+x_{d-1}$.

\begin{figure}[h!]\centering
	\includegraphics[width=\linewidth]{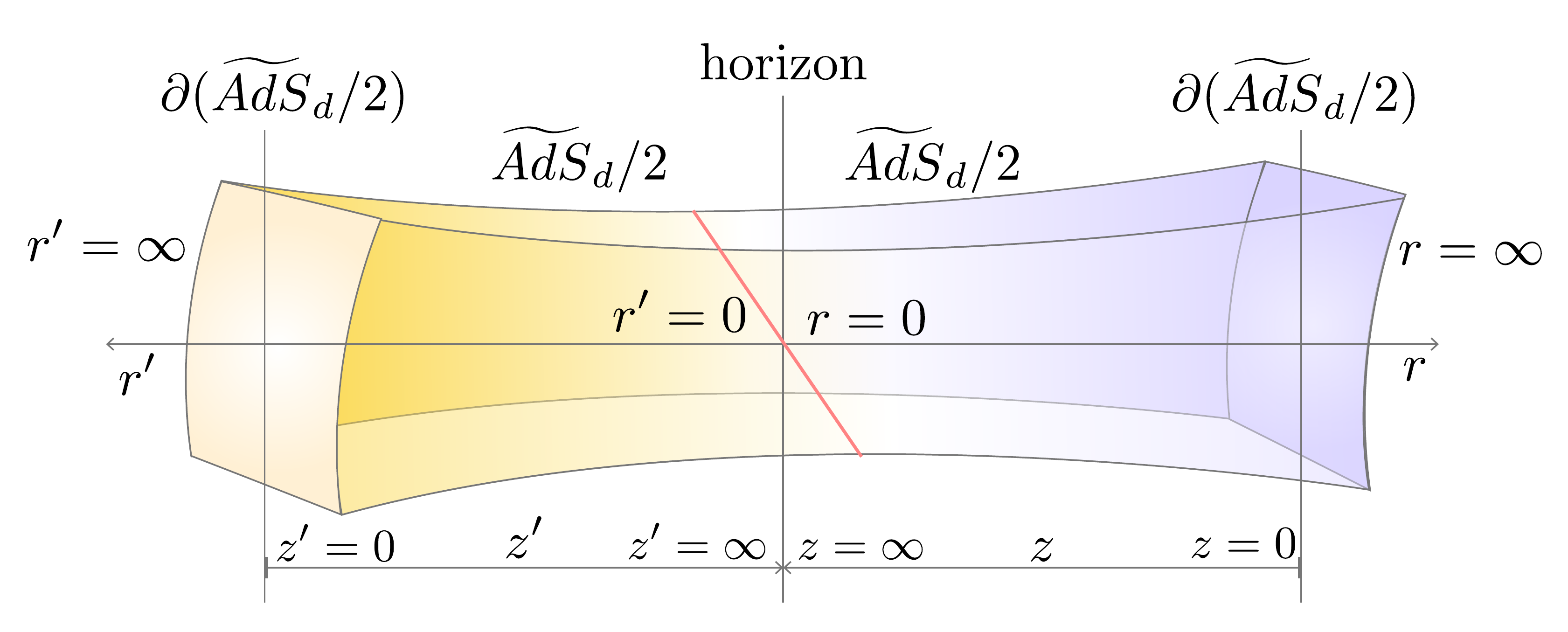}
	\caption{Two patches of $ \widetilde{AdS_d} $}
\end{figure}

Consider e.g. $\widetilde{AdS_3}\subset\mathbb{E}^{(2,2)}$, with metric $$ds^2_{\widetilde{AdS_3}}={{a^2}\over{cos^2\chi}}({dt^\prime}^2-d\chi^2-sin^2\chi d\varphi^2).\eqno{(3.6b)}$$ The spacetime is a solid infinite cylinder without boundary of radius $\pi/2$ (see Fig. 6). From (3.17a) one sees that, at $r$, the Newtonian approximation to the gravitational force is $-r/a^2$ which is attractive towards $r=0$ ($\chi=0$) and that the proper time interval is $\Delta\tau(r)=\sqrt{1+{{r^2}\over{a^2}}}\Delta T$. Then at the origin $\Delta\tau$ coincides with the coordinate time interval. 

\

3.11.c {\it Geodesics}

\

The equation of motion of radial light rays (mass $m=0$) is $d\chi=dt^\prime$; then $\Delta\tau=\Delta t^\prime$ and therefore the proper time measured at the origin for a light ray which goes to and bounces off infinity ($r\to+\infty$ or $\chi\to\pi/2_-$)  is finite and equals $2(a\pi/2)=\pi a$.

\begin{figure}[h!]\centering
	\includegraphics[width=.7\linewidth]{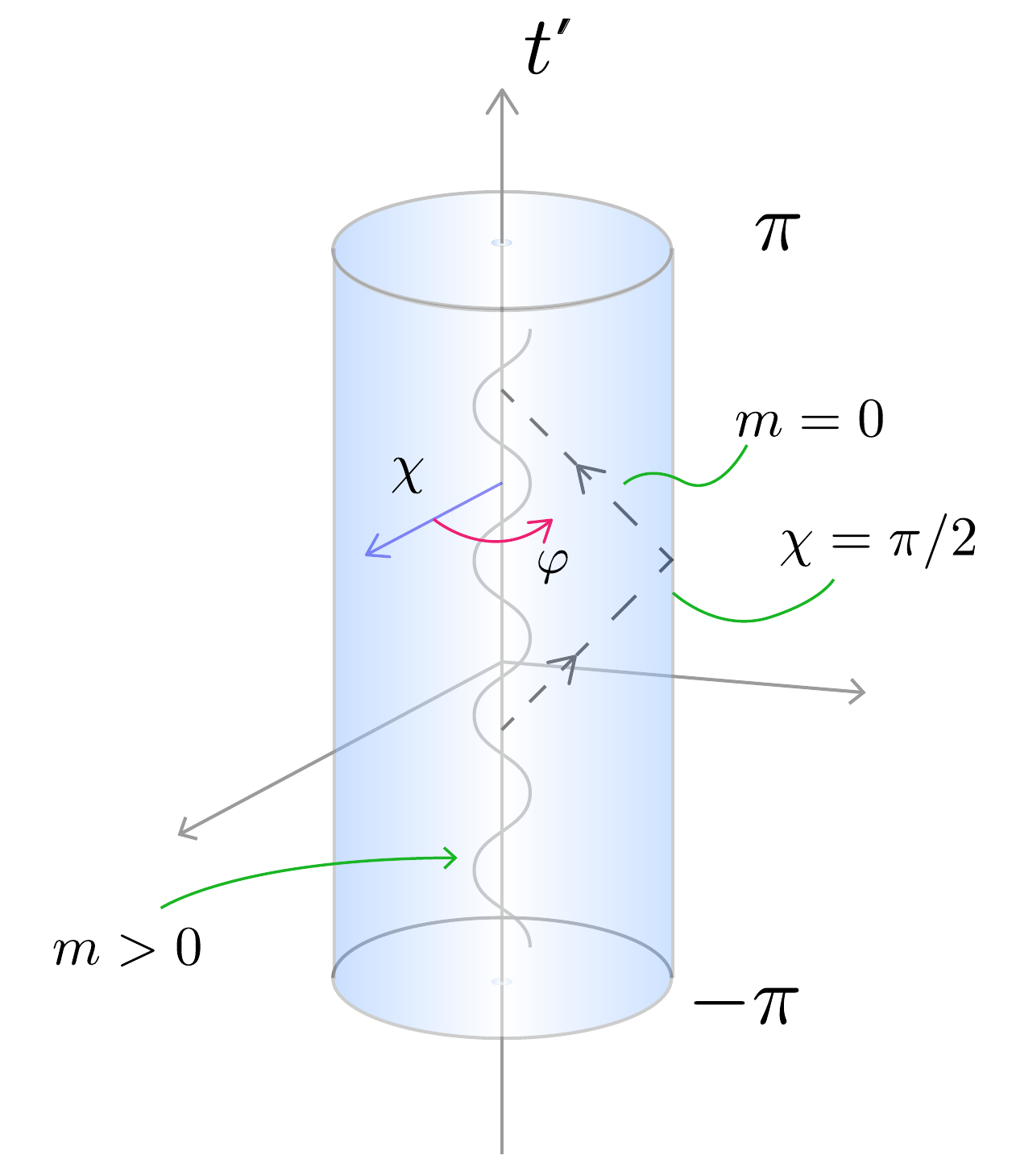}
	\caption{Radial massive and massless geodesics in $ \widetilde{AdS_3} $}
\end{figure}

On the other hand, for timelike radial geodesics (mass $m>0$), $$1={({{d}\over{d\tau}}s_{\widetilde{AdS_3}})}^2\equiv {\cal L}:={{a^2}\over{cos^2\chi}}(({\dot{t}}^\prime)^2-{\dot{\chi}}^2),\eqno{(3.39)}$$ where ${\dot{t}}^\prime={{dt^\prime}\over{d\tau}}$ and $\dot{\chi}={{d\chi}\over{d\tau}}$. From the Lagrange equation ${{d}\over{d\tau}}({{\partial {\cal L}}\over{\partial{\dot{t}}^\prime}})={{\partial {\cal L}}\over{\partial t^\prime}}=0$ one obtains ${{d}\over{d\tau}}({{{\dot{t}}^\prime}\over{cos^2\chi}})=0$ i.e. $${\dot{t}}^\prime=k \ cos^2\chi\eqno{(3.40)}$$ with $k$=const., $[k]=L^{-1}$. Replacing (3.40) in (3.39) one obtains $${{d\chi}\over{dt^\prime}}=\pm\sqrt{1-{{1}\over{k^2a^2cos^2\chi}}}\eqno{(3.41)}$$ which implies $\vert k\vert\geq 1/a$. The integration of (3.41) leads to the periodic solution $$\chi(t^\prime)=arcsin(\sqrt{1-{{1}\over{\vert ka\vert^2}}}sin(t^\prime-t_0^\prime))=\chi(t^\prime+2\pi n).\eqno(3.42)$$ $\chi(t^\prime)<\pi/2$ for all $t^\prime$ i.e. the particle never reaches the boundary. In particular, for $\vert k\vert a=1$ i.e. $\vert k\vert =a^{-1}$, the particle remains at rest at the spatial origin $\chi =0$ since, from (3.42), $\chi (t^\prime)=0$ for all $t^\prime$. 

\

3.12. {\it $\widetilde{AdS_4}$ as a stack of $H^3$'s}

\

From (3.6) and (2.35) one sees that, at each $t^\prime\in(-\infty,+\infty)$, $$\widetilde{AdS_4}\vert_{t^\prime}=H^3.\eqno{(3.43)}$$ This holds for each of the two patches of $\widetilde{AdS_4}$.

\
 
This fact generalizes to the $d$-dimensional anti-De Sitter spacetime. The metric of the ($d-1$)-dimensional hyperbolic space is $$dl^2_{H^{d-1}}={{a^2}\over{cos^2\chi}}(d\chi^2+sin^2\chi \ d\Omega^2_{d-2}).\eqno{(3.44)}$$ Comparing with (3.6a), it is clear that $\widetilde{AdS_d}$ is a stack of $H^{d-1}$'s i.e. $$\widetilde{AdS_d}\vert_{t^\prime}=H^{d-1}.\eqno{(3.45)}$$ In particular $$\widetilde{AdS_5}\vert_{t^\prime}=H^4.\eqno{(3.46)}$$

\

3.13. {\it Tortoise radial coordinate}

\

We conclude Section 3 defining the tortoise radial coordinate $r^*$ for the $\widetilde{AdS_4}$ case, in preparation for its definition in the $S\widetilde{AdS_4}$ in Subsection 4.5.

\

(3.17) can be written in the form $$ds^2_{\widetilde{AdS}_4}=(1+{{r^2}\over{a^2}})(dt^2-{dr^*}^2)-r^2d\Omega_2^2,\eqno{(3.47)}$$ where $$dr^*:={{dr}\over{1+{{r^2}\over{a^2}}}}.\eqno{(3.48)}$$ Integrating one obtains the {\it Regge-Wheeler} or {\it tortoise} radial coordinate: $$r^*(r)=\int_0^r{{dr^\prime}\over{1+{{{r^\prime}^2}\over{a^2}}}}=a \ arctg({{r}\over{a}}),\eqno{(3.49)}$$ with inverse $r(r^*)=a \ tg({{r^*}\over{a}})$. In particular, $r^*(0)=0$, $r^*(a)={{\pi}\over{4}}a$, $r^*(\infty)={{\pi}\over{2}}a$. (See Fig. 7.) In this coordinate, radial light rays move in the $t/r^*$ plane at $45^o$ or $135^o$: $$({{dt}\over{dr^*}})^2=1 \ \Rightarrow \ {{dt}\over{dr^*}}=\pm 1. \eqno{(3.50)}$$

\begin{figure}[h!]\centering
	\includegraphics[width=.7\linewidth]{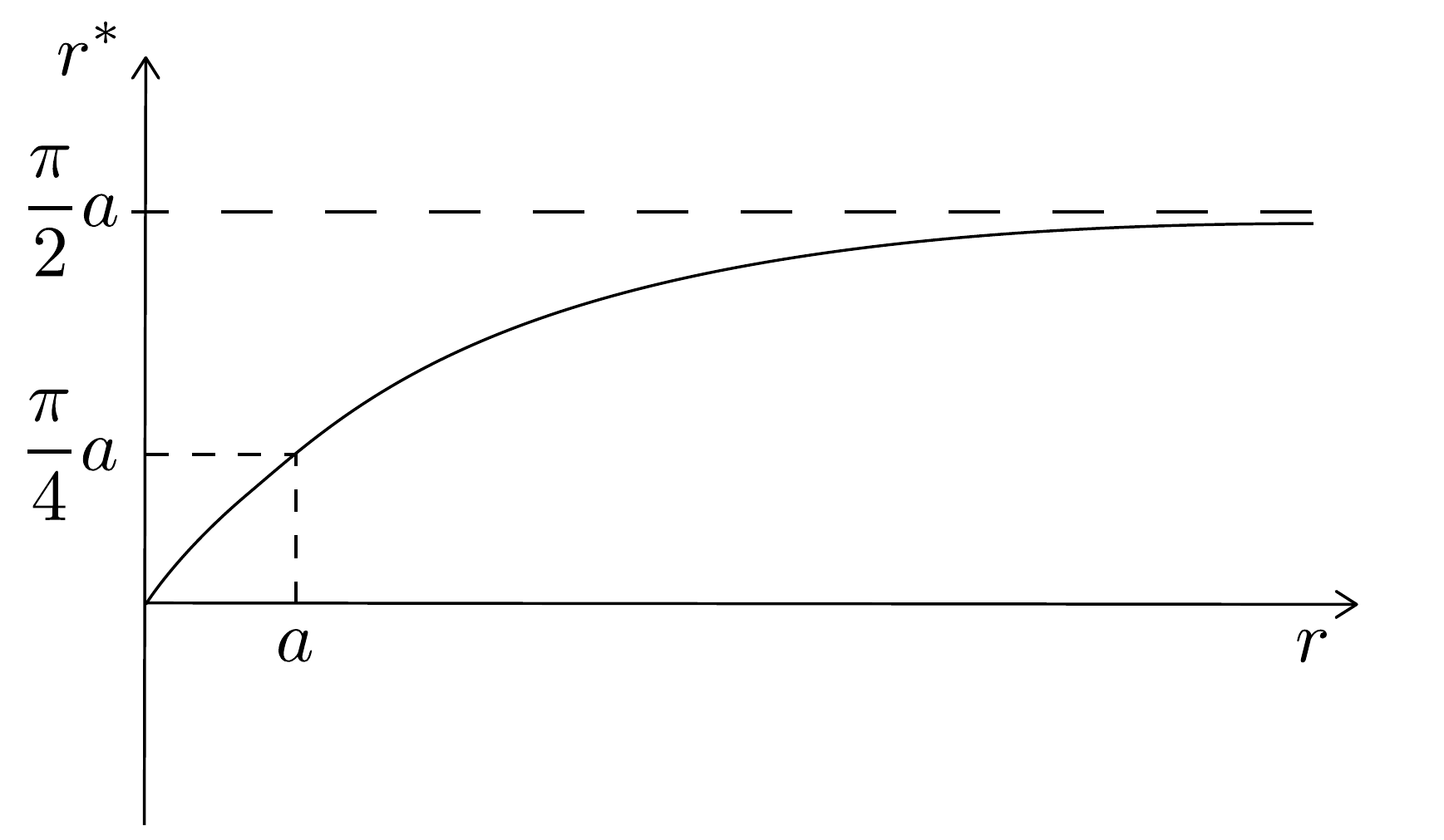}
	\caption{Tortoise coordinate for $ \widetilde{AdS_4} $}
\end{figure}

\section{Four dimensional Schwarzschild anti-De Sitter metric ($S\widetilde{AdS}_4$)}

\

4.1.{\it Schwarzschild coordinates}

\

The general form of a static spherically symmetric metric is $$ds^2=e^{2\nu}dt^2-e^{2\lambda}dr^2-r^2d\Omega^2_2\eqno{(4.1)}$$ where $\nu=\nu(r)$, $\lambda=\lambda(r)$. We'll call $(x^0,x^1,x^2,x^3)=(t,r,\theta,\varphi)$, with $t\in(-\infty,+\infty)$, $r\in[0,+\infty)$, $\theta\in[0,\pi]$, $\varphi\in[0,2\pi)$; $[t]=[r]=L^1$, $[\theta]=[\varphi]=L^0$. For the metric tensor one has $$diag(g_{00},g_{11},g_{22},g_{33})=(e^{2\nu},-e^{2\lambda},-r^2,-r^2sin^2\theta),\eqno{(4.2)}$$ with inverse $${g_{\mu\nu}}^{-1}\equiv g^{\mu\nu}=diag(g^{00},g^{11},g^{22},g^{33})=(e^{-2\nu},-e^{-2\lambda},-r^{-2},-r^{-2}sin^{-2}\theta).\eqno{(4.3)}$$ The metric is parity ($\theta\to\pi-\theta, \ \varphi\to \varphi+\pi$) and time reversal ($t\to -t$) invariant, and has two Killing vector fields: $\partial_t$ and $\partial_\varphi$. 

\

There are nine algebraically independent non vanishing Christoffel symbols (2.25): $$\Gamma^0_{01}=\nu^\prime, \ \ \Gamma^2_{12}=\Gamma^3_{13}={{1}\over{r}},$$ $$\Gamma^1_{11}=\lambda^\prime, \ \ \Gamma^2_{33}=-sin\theta cos\theta,$$ $$\Gamma^{1}_{22}=-re^{-2\lambda}, \ \ \Gamma^3_{23}=cotg\theta,$$ $$\Gamma^1_{33}=-re^{-2\lambda}sin^2\theta, \ \ \Gamma^1_{00}=\nu^\prime e^{2(\nu-\lambda)}\eqno{(4.4)}$$ where $\nu^\prime={{d\nu}\over{dr}}$ and $\lambda^\prime={{d\lambda}\over{dr}}.$ For the Ricci tensor $$R_{\sigma\nu}=R^\rho_{\sigma\rho\nu}=\Gamma^\rho_{\sigma\nu,\rho}-\Gamma^\rho_{\sigma\rho,\nu}+\Gamma^\lambda_{\sigma\nu}\Gamma^\rho_{\lambda\rho}-\Gamma^\lambda_{\sigma\rho}\Gamma^\rho_{\lambda\nu}\eqno{(4.5)}$$ one obtains the four non vanishing components $$R_{00}=(\nu^{\prime\prime}-\nu^\prime\lambda^\prime+{\nu^\prime}^2+{{2\nu^\prime}\over{r}})e^{2(\nu-\lambda)},\eqno{(4.6a)}$$ $$R_{11}=-\nu^{\prime\prime}+{{2\lambda^\prime}\over{r}}-{\nu^\prime}^2+\lambda^\prime\nu^\prime,\eqno{(4.6b)}$$ $$R_{22}=(r\lambda^\prime-r\nu^\prime-1)e^{-2\lambda}+1,\eqno{(4.6c)}$$ $$R_{33}=sin^2\theta R_{22}.\eqno{(4.6d)}$$ The vacuum Einstein equations with cosmological constant $\Lambda$ are $$R_{\mu\nu}-{{1}\over{2}}g_{\mu\nu}R-\Lambda g_{\mu\nu}=0.\eqno{(4.7)}$$ Contracting indices and using again (4.7) leads to $$R_{\mu\nu}=-\Lambda g_{\mu\nu}.\eqno{(4.8)}$$ From (4.6a,b), (4.8) and (4.2), $$(\nu^{\prime\prime}-\lambda^\prime\nu^\prime+{\nu^\prime}^2+{{2\nu^\prime}\over{r}})e^{2(\nu-\lambda)}=-\Lambda e^{2\nu},$$ $$-\nu^{\prime\prime}+\lambda^\prime\nu^\prime-{\nu^\prime}^2+{{2\lambda^\prime}\over{r}}=\Lambda e^{2\lambda}$$ which imply $$\nu^{\prime\prime}-\lambda^\prime\nu^\prime+{\nu^\prime}^2+{{2\nu^\prime}\over{r}}=-\Lambda e^{2\lambda},$$ $$-\nu^{\prime\prime}+\lambda^\prime\nu^\prime-{\nu^\prime}^2+{{2\lambda^\prime}\over{r}}=\Lambda e^{2\lambda}$$ and therefore $$\nu^\prime+\lambda^\prime=0 \ \ i.e. \ \ \lambda(r)=-\nu(r)+const.\eqno{(4.9)}$$ In the asymptotically flat case (Schwarzschild), $\lambda,\nu\to 0$ as $r\to\infty$, which requires $const.=0$; for the $\widetilde{AdS}_4$ case, from (3.17), $e^{2\nu}=1+{{r^2}\over{a^2}}$ and $e^{2\lambda}={{1}\over{1+{{r^2}\over{a^2}}}}$; then $\nu={{1}\over{2}}ln(1+{{r^2}\over{a^2}})\to ln({{r}\over{a}})$ and $\lambda=-{{1}\over{2}}ln(1+{{r^2}\over{a^2}})\to -ln({{r}\over{a}})$ as $r\to+\infty$ i.e. $\nu+\lambda\to 0$ in this limit and therefore $const.=0$, consistent with the previous case. Then, $$\lambda(r)=-\nu(r).\eqno{(4.9a)}$$ (4.6d) holds automatically. From (4.6c) and (4.9a), $(1+2r\nu^\prime)e^{2\nu}-1=-\Lambda r^2$ i.e. ${{d}\over{dr}}(re^{2\nu})=1-\Lambda r^2$ which implies $e^{2\nu}=1+{{const.}\over{r}}-{{\Lambda r^2}\over{3}}$ i.e. $g_{00}=1+{{const.}\over{r}}-{{\Lambda r^2}\over{3}}.$ 

\

$r=0$ is the {\it curvature singularity} i.e. where scalars (and therefore coordinate invariants) constructed with the curvature tensor $R^{\mu}_{\nu\rho\sigma}$ (e.g. $R_{\mu\nu\rho\sigma}R^{\mu\nu\rho\sigma}$) diverge.

\

For small $r$, the Newtonian approximation gives $const.=2M$, where $M$ is the gravitational mass; so, $$g_{00}={(g_{11})}^{-1}=1+{{2M}\over{r}}-{{\Lambda r^2}\over{3}}\eqno{(4.10)}$$ and therefore $$ds^2_{S\widetilde{AdS}_4}=(1-{{2M}\over{r}}+{{r^2}\over{a^2}})dt^2-{{dr^2}\over{1-{{2M}\over{r}}+{{r^2}\over{a^2}}}}-r^2d\Omega^2_2,\eqno{(4.11)}$$ where we used the relation (3.20) between the cosmological constant and the curvature radius of the $AdS$ spacetime. This is the starting point of the study of the $S\widetilde{AdS}_4$ black hole. 

\

4.2. {\it Horizon}

\

The horizons are given by the zeros of $g_{00}$ i.e. the real roots of $$f(r):=1-{{2M}\over{r}}+{{r^2}\over{a^2}},\eqno{(4.12)}$$ or, equivalently $$r^3+a^2r-2Ma^2=0.\eqno{(4.13)}$$ $f(r)\to -\infty$ as $r\to 0_+$ and $f(r)\to +\infty$ as $r\to +\infty$. It is easy to check that $f(r)$ has: i) no extrema: $f^\prime(r)={{2M}\over{r^2}}+{{2r}\over{a^2}}=0\Rightarrow r^3=-Ma^2<0$, and ii) one inflection point: $f^{\prime\prime}(r)=-{{4M}\over{r^3}}+{{2}\over{a^2}}=0\Rightarrow r=\bar{r}=(2Ma^2)^{1/3}$. Then the spacetime ${S\widetilde{AdS}_4}$ has only one horizon $r_h=r_h(2M,a)$. (The other two roots of $f(r)$ are complex and have no physical meaning.) (See Fig. 8.)

\begin{figure}[h!]\centering
	\includegraphics[width=.5\linewidth]{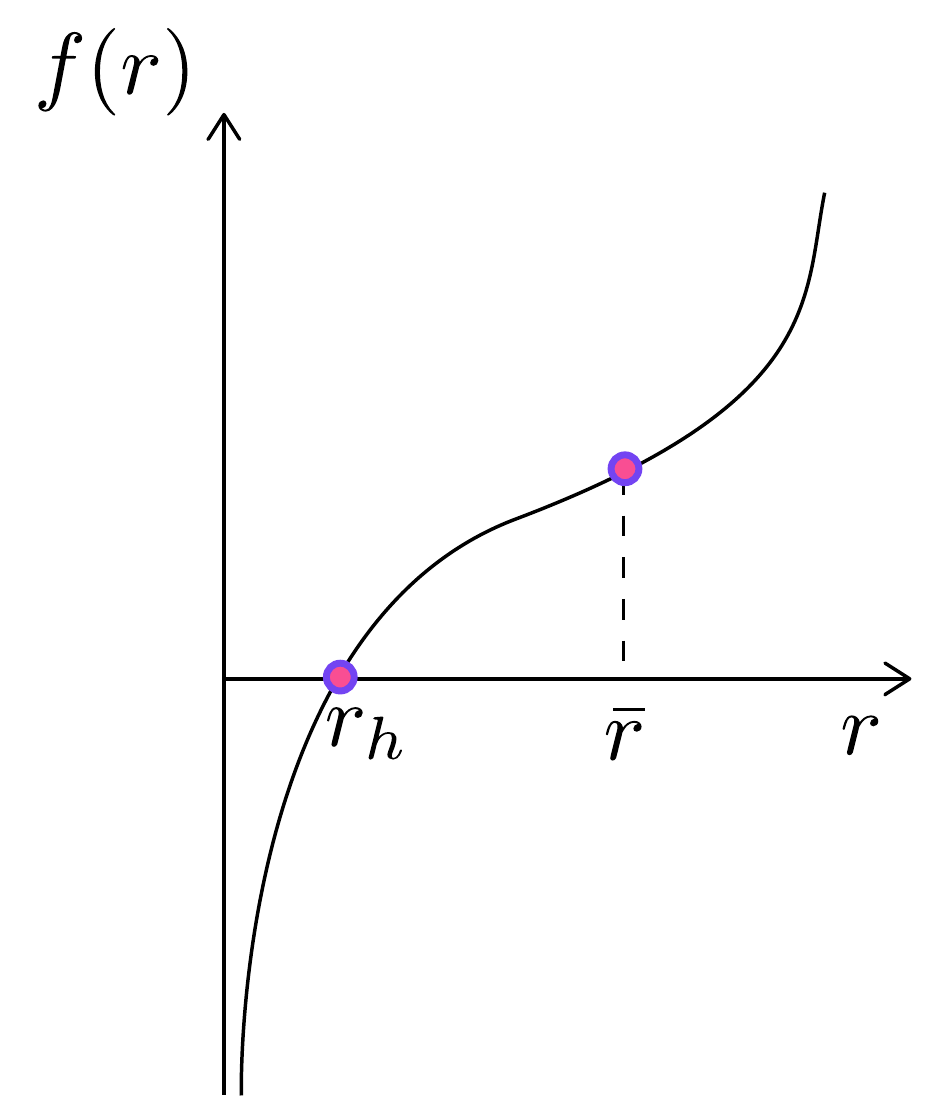}
	\caption{Horizon for $ \widetilde{SAdS_4} $}
\end{figure}

For a solar mass, $2M\sim 10^3m$; on the other hand, $\Lambda=-{{3}\over{a^2}}\sim -10^{-60}m^{-2}$ i.e. $a^2\sim 10^{60}m^2$; so, typically $2M<<a$ i.e. ${{M}\over{a}}\cong 10^{-27}<<1$. For the position of the inflection point one has $\bar{r}\sim {(10^{63}m^3)}^{1/3}=10^{21}m=10^{18}km\sim 10^7 lyrs$.  

\

If $q^2/4+p^3/27>0$ then the polynomial $r^3+pr+q$ has the real root $(-{{q}\over{2}}+\sqrt{{{q^2}\over{4}}+{{p^3}\over{27}}})^{1/3}+(-{{q}\over{2}}-\sqrt{{{q^2}\over{4}}+{{p^3}\over{27}}})^{1/3}$; this condition holds in our case: $q=-2Ma^2$, $p=a^2\Rightarrow q^2/4+p^3/27=a^4(M^2+a^2/27)>0.$ So, the {\it horizon for the} $S\widetilde{AdS}_4$ metric is $$r_h=r_h(M,a)=(Ma^2)^{1/3}((1+\sqrt{1+{{a^2}\over{27M^2}}})^{1/3}+(1-\sqrt{1+{{a^2}\over{27M^2}}})^{1/3}).\eqno{(4.14)}$$ 

\

From (4.12) one also obtains, at $r_h$, $$M={{r_h}\over{2}}(1+{{r_h^2}\over{a^2}}), \eqno{(4.15)}$$ which gives $M=M(r_h,a)$, a sort of inverse of (4.14).

\

It is interesting to compute the deviation of $r_h$ from $2M=r_{Schw.}$ for the case $M/a<<1$. From (4.15), ${{M}\over{a}}={{1}\over{2}}({{r_h}\over{a}})(1+({{r_h}\over{a}})^2)<<1$ implies ${{r_h}\over{a}}\simeq{{M}\over{a}}<<1$ and therefore $$r_h={{2M}\over{1+{{r_h^2}\over{a^2}}}}\simeq 2M(1-({{r_h}\over{a}})^2)=r_{Schw.}(1-({{r_h}\over{a}})^2)=r_{Schw.}(1-O({{M}\over{a}})^2). \eqno{(4.16)}$$

\

 Clearly, $r_h\to (r_{Schw.})_-$ as $a\to+\infty$ ($\Lambda\to 0_-$), $r_h\to 0$ as $M\to 0$ (case of $\widetilde{AdS}_4$), and $r_h\to(2Ma^2)^{1/3}$ as $M\to\infty$. Also, $\bar{r}=r_h(1+({{a}\over{r_h}})^2)^{1/3}>>r_h$ for ${{a}\over{M}}>>1$. 

\

4.3. {\it Surface gravity}

\

The {\it surface gravity} ($\kappa$) of a black hole is the magnitude $A$ of the 4-acceleration of a static observer at the horizon ($r_h$) as measured by a static observer at infinity. A static observer at the horizon must be accelerated: on the contrary its motion would be geodesic i.e. in free fall. The 4-velocity of such an observer at $r>r_h$ is $u^\mu_{obs.}(r)={{dx^\mu}\over{d\tau}}$ with $\vec{u}=\vec{0}$. So, $u^0_{obs.}(r)={{dx^0}\over{d\tau}}={{dt}\over{(1-{{2M}\over{r}}+{{r^2}\over{a^2}})^{1/2}dt}}=(1-{{2M}\over{r}}+{{r^2}\over{a^2}})^{-1/2}$ and therefore $$u^\mu_{obs.}(r)=(u^0_{obs.}(r),\vec{0})=((1-{{2M}\over{r}}+{{r^2}\over{a^2}})^{-1/2},\vec{0}).\eqno{(4.17)}$$ In particular, $u^0_{obs.}(r)\to +\infty$ as $r\to (r_h)_+$, and $u^0_{obs.}(r)\to 1$ as $r\to \bar{r}$. 

\

For the computation of $\kappa$ we need the explicit expressions for the Christoffel symbols. From (4.2), (4.4), (4.9a) and (4.10) the result is: $$\Gamma^0_{01}=-\Gamma^1_{11}={{{{M}\over{r^2}}+{{r}\over{a^2}}}\over{1-{{2M}\over{r}}+{{r^2}\over{a^2}}}},$$ $$\Gamma^1_{22}=-r+2M-{{r^3}\over{a^2}},$$ $$\Gamma^1_{33}=(-r+2M-{{r^3}\over{a^2}})sin^2\theta,$$ $$\Gamma^2_{12}=\Gamma^3_{13}={{1}\over{r}}, \ \Gamma^2_{33}=-sin\theta \ cos\theta, \ \Gamma^3_{23}=cotg\theta,$$ $$\Gamma^1_{00}=(1-{{2M}\over{r}}+{{r^2}\over{a^2}})({{M}\over{r^2}}+{{r}\over{a^2}}). \eqno{(4.18)}$$ For the 4-acceleration, the covariant derivative of the 4-velocity, namely $$a^\mu={{Du^\mu}\over{d\tau}}={{du^\mu}\over{d\tau}}+\Gamma^\mu_{\nu\rho}u^\nu u^\rho,\eqno{(4.19)}$$ one obtains $$a^\mu=(0,{{M}\over{r^2}}+{{r}\over{a^2}},0,0), \ a_\mu=g_{\mu\nu}a^\nu=-({{M}\over{r^2}}+{{r}\over{a^2}})/(1-{{2M}\over{r}}+{{r^2}\over{a^2}}), \eqno{(4.20)}$$ and therefore $$A=A(r)=\sqrt{-a^\mu a_\mu}=({{M}\over{r^2}}+{{r}\over{a^2}})/(1-{{2M}\over{r}}+{{r^2}\over{a^2}})^{1/2}.\eqno{(4.21)}$$ We notice that $A(r)\to +\infty$ as $r\to (r_h)_+$: to maintain an observer at rest at the horizon requires an infinite acceleration. Its red-shift at infinity is however finite and is given by $$A_\infty(r):=(1-{{2M}\over{r}}+{{r^2}\over{a^2}})^{1/2}A(r)={{M}\over{r^2}}+{{r}\over{a^2}}.\eqno{(4.22)}$$ The surface gravity is its value at $r_h$: $$\kappa=A_\infty(r_h)={{M}\over{(r_h(M,a))^2}}+{{r_h(M,a)}\over{a^2}}=\kappa(M,a).\eqno{(4.23)}$$ For $a=+\infty$, $r_h=2M$ and $A_\infty(2M)={{1}\over{4M}}=\kappa_{Schw.}$. 

\

For $M\to 0$, $r_h\to 2M(1-\sqrt{3}{{M}\over{a}})\to 0$, and $\Rightarrow\kappa\to{{1}\over{4M}}\to +\infty$; for $M\to +\infty$, $r_h\to (2Ma^2)^{1/3}$ and $\Rightarrow\kappa\to 3({{M}\over{4a^4}})^{1/3}\to +\infty$. So, $\kappa$ has a minimum at $M=M_0$ determined by the condition ${{\partial\kappa}\over{\partial M}}\vert_{M_0}=0$ i.e. $$({{2M}\over{r_h}}-{{r_h^2}\over{a^2}})\vert_{M_0}{{\partial r_h}\over{\partial M}}\vert_{M_0}=1,\eqno{(4.24)}$$ with $M_0=M_0(a)$. Using (4.16), for the minimum one obtains $$r_h={{a}\over{\sqrt{3}}}\eqno{(4.25)}$$ and $$M_0(a)={{2a}\over{3\sqrt{3}}}.\eqno{(4.26)}$$ (See Fig. 9.)

\begin{figure}[h!]\centering
	\includegraphics[width=.7\linewidth]{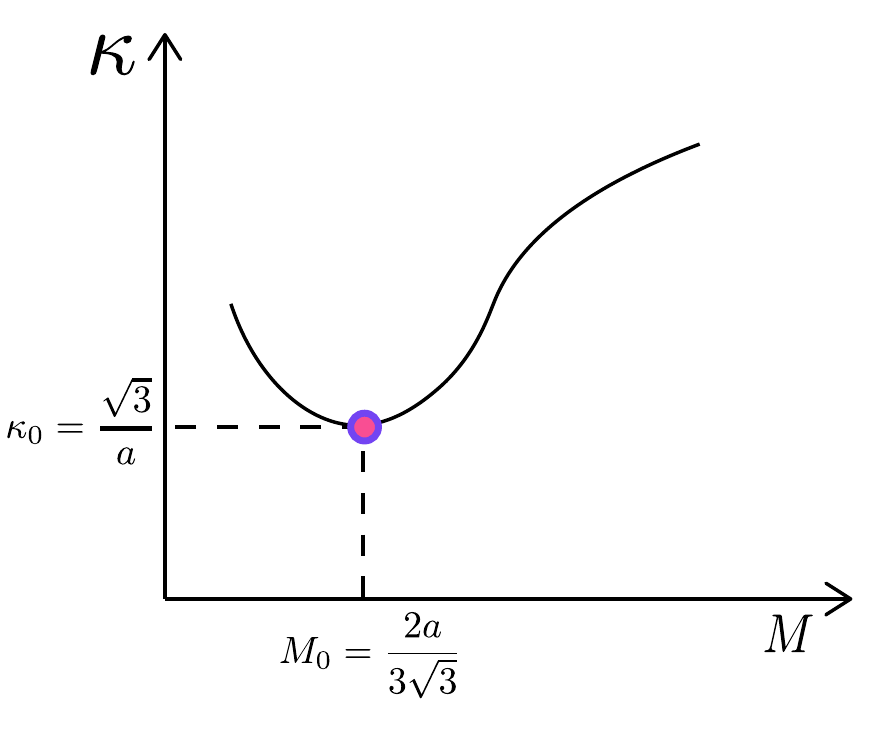}
	\caption{Surface gravity as function of mass for $ \widetilde{SAdS_4} $}
\end{figure}

4.4. {\it Rindler approximation and Hawking temperature}

\

We study the time-radial ($t/r$) part of the $S\widetilde{AdS}_4$ metric near the horizon defining the coordinate $\rho$, with $\vert\rho\vert<<r_h$, $[\rho]=L^1$, through $$r=r_h+{{\alpha\rho^2}\over{r_h}},\eqno{(4.27)}$$ where $\alpha\in\mathbb{R}$ to be determined later, and keeping terms up to $O(\rho^2)$. A straightforward calculation from (4.11) leads to $$ds^2_{S\widetilde{AdS}_4}\vert_{t/r}={{2\alpha}\over{r_h\kappa}}((\kappa\rho)^2dt^2-d\rho^2)\eqno{(4.28)}$$ which is conformal to the Rindler metric [10] $$ds^2_R=(a_R\rho)^2-d\rho^2\eqno{(4.29)}$$ with Rindler acceleration $$a_R=\kappa.\eqno{(4.30)}$$ With the choice $\alpha={{r_h\kappa}\over{2}}$, $$ds^2_{S\widetilde{AdS}_4}\vert_{t/r}=ds^2_R,\eqno{(4.31)}$$ and the coordinate transformation (4.27) is $$r=r_h+{{\kappa}\over{2}}\rho^2.\eqno{(4.32)}$$ By the Unruh effect, which basically consists in the appearance of thermal radiation for any uniformly accelerated observer in the Minkowski vacuum [11], the Rindler temperature $$T_R={{a_R}\over{2\pi}}\eqno{(4.33)}$$ can be identified with the Hawking temperature at the horizon $r_h$ of the $S\widetilde{AdS}_4$ black hole: $$T_{Hawk.}\vert_{S\widetilde{AdS}_4}={{\kappa}\over{2\pi}}={{1}\over{2\pi}}({{M}\over{r_h^2}}+{{r_h}\over{a^2}})={{3r_h^2+a^2}\over{4\pi a^2r_h}}\eqno{(4.34)}$$ where in the last equality we used (4.15). This is the temperature at which the black hole exists in stable thermodynamic equilibrium with thermal radiation produced by quantum vacuum fluctuations near the horizon.

\

Due to (4.24) $T_{Hawk.}\vert_{S\widetilde{AdS}_4}\equiv T_H$ attains a minimum $T_0$ at $M_0(a)$ given by $$T_0={{\sqrt{3}}\over{2\pi a}}\eqno{(4.35)}$$ i.e. for $\kappa=\kappa_0={{\sqrt{3}}\over{a}}$. For any $T_H>T_0$ there are two black hole solutions: the smaller one, with $M<M_0(a)$, has negative specific heat: ${{\partial M}\over{\partial T_H}}<0$ (temperature decreasing with increasing mass), and therefore is thermodinamically unstable, while the larger one, with $M>M_0(a)$, has positive specific heat and is thermodinamically stable. For $T_H<T_0$ there is no black hole.

\

4.5. {\it Kruskal-Szekeres coordinates}

\

The time/radial part of the metric (4.11) can be written in the form $$ds^2_{S\widetilde{AdS}_4}\vert_{t/r}=(1-{{2M}\over{r}}+{{r^2}\over{a^2}})(dt^2-{dr*}^2)\eqno{(4.36)}$$ where the tortoise coordinate $r^*$ satisfies the equation $${dr^*}^2={{dr^2}\over{(1-{{2M}\over{r}}+{{r^2}\over{a^2}})^2}}.\eqno{(4.37)}$$ This can be integrated, with the result $$r^*(r)=\int_0^r{{dr^\prime}\over{1-{{2M}\over{r^\prime}}+{{{r^\prime}^2}\over{a^2}}}}={{a^2}\over{3r_h^2+a^2}}(r_hln\vert1-{{r}\over{r_h}}\vert-{{r_h}\over{2}}ln(1+{{r(r+r_h)}\over{r_h^2+a^2}})$$ $$+{{3r_h^2+2a^2}\over{\sqrt{3r_h^2+4a^2}}}arctg({{r\sqrt{3r_h^2+4a^2}}\over{2(r_h^2+a^2)+rr_h}})).\eqno{(4.38)}$$ $r^*(r)$ satisfies: $$r^*(r_h)=-\infty, \ r^*(0)=0, \ r^*(+\infty)={{a^2}\over{3r_h^2+a^2}}(r_hln\sqrt{1+({{a}\over{r_h}})^2}+{{3r_h^2+2a^2}\over{\sqrt{3r_h^2+4a^2}}}arctg({{\sqrt{3r_h^2+4a^2}}\over{r_h}})).\eqno{(4.39)}$$ Also, $r^*(r)\to a \ arctg({{r}\over{a}})$ as $M\to 0$ (case of $\widetilde{AdS}_4$).

\

Strictly speaking, one should divide the domain of integration from $r=0$ to $r_h-\varepsilon$ and from $r_h+\varepsilon$ to $r>r_h$. In this case the resulting functions $r^*(r)$ after taking the limit $\varepsilon\to 0_+$ would differ by an irrelevant constant. 

\

A careful analysis of (4.38) shows that the behavior of $r^*$ with $r$ is that shown in Fig. 10. First of all, the divergence of $r^*$ as $r\to r_{h}$ reflects the pole in the integrand of the integral in (4.38) at $r=r_{h}$; secondly, $r^*(r)$ exhibits two branches: for $0\leq r<r_h$, $r^*$ is monotonously decreasing, while for $r_h<r$, $r^*$ is monotonously increasing; given $0<r^*<r^*(\infty)$ the solution for $r$ is clearly unique, while for $r^*<0$ there are two solutions for $r$: one for $r<r_h$ and another for $r_h<r$. In each region of the maximally extended spacetime diagram, $r^*(r)$ can then be inverted. $\hat{r}$ is such that $r^*(\hat{r})=0$.

\begin{figure}[h!]\centering
	\includegraphics[width=.7\linewidth]{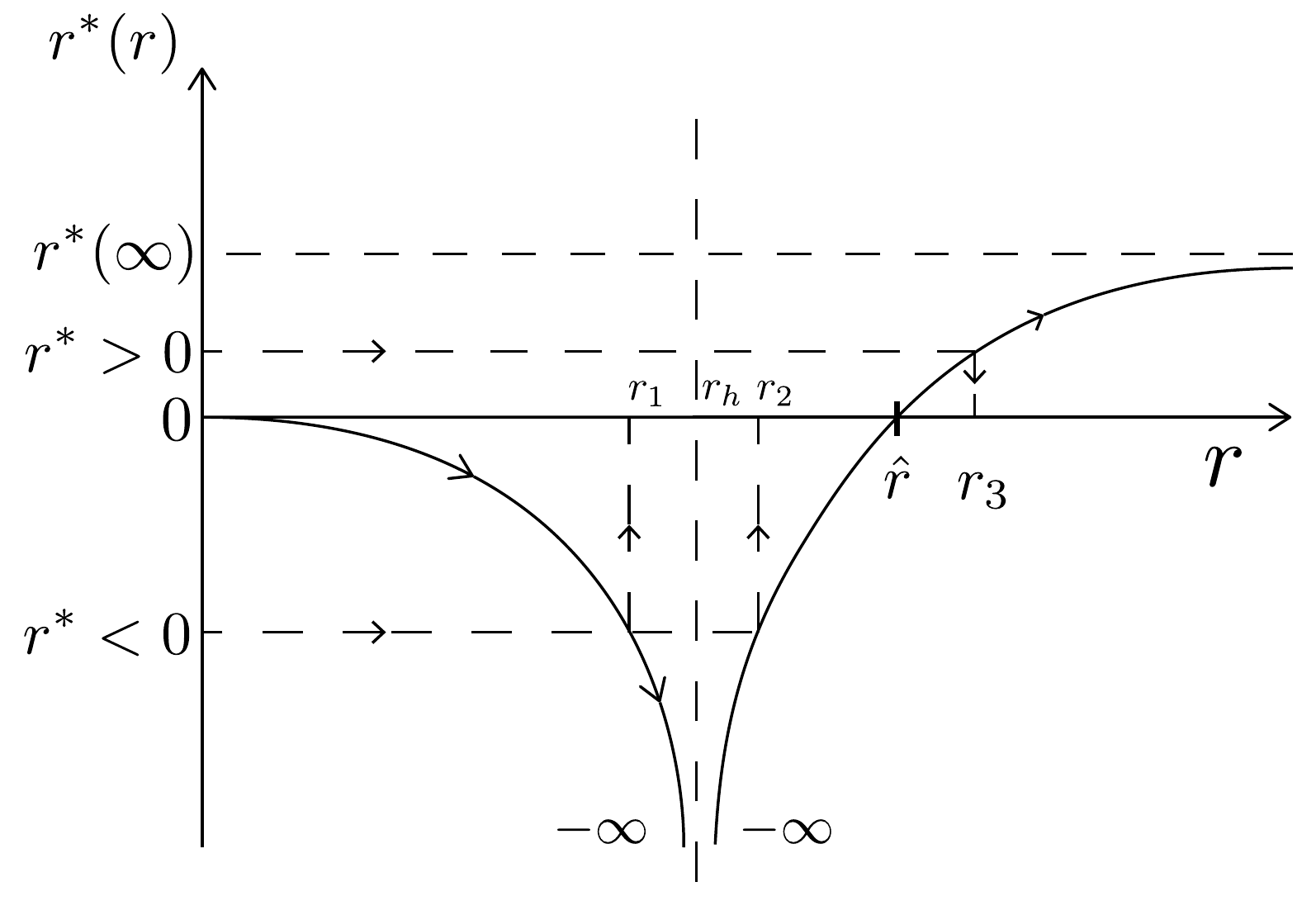}
	\caption{Tortoise coordinate $ r^* $ for $ \widetilde{SAdS_4} $}
\end{figure}

We define the dimensionless coordinates $$U:=-sgn(f)e^{f^\prime(r_h){{r^*(r)-t}\over{2}}}, \ V:=e^{f^\prime(r_h){{r^*(r)+t}\over{2}}},\eqno{(4.40)}$$ where $f$ is given by (4.12) and therefore $$U<0 \ for \ r>r_h, \ U>0 \ for \ r<r_h, \ v>0 \ for \ all \ r, \ f^\prime(r_h)={{df(r)}\over{dr}}\vert_{r=r_h}=2({{M}\over{r_h^2}}+{{r_h}\over{a^2}})=4\pi T_{Hawk.}\vert_{S\widetilde{AdS}_4},\eqno{(4.41)}$$ and, since $r^*(r_h)=-\infty$, $$U(r_h)=0 \ (+V-axis), \ V(r_h)=0 \ (U-axis).\eqno{(4.42)}$$ The axis $U$ and $V$ are $135^o$ and $45^o$ with respect to the $X$-axis to be defined in eq. (4.50) (Fig.11). (4.40) covers the regions $(I)$ ($r>r_h$) and $(III)$ ($r<r_h$)in the whole $U/V$ plane. 

\

From (4.40), $$t=-{{1}\over{f^\prime(r_h)}}ln(-sgn(f)({{U}\over{V}})).\eqno{(4.43)}$$ Therefore: 

\

i) $V\to 0_+\Rightarrow t\to -\infty$ for both $U>0$ ($+U$-axis, $r<r_h$) and $U<0$ ($-U$-axis, $r>r_h$); ii) $t\to +\infty$ for both $U\to 0_+$ ($+V$-axis, $r<r_h$) and $U\to 0_-$ ($+V$-axis, $r>r_h$); iii) $V=U>0 \Rightarrow sgn(f)<0\Rightarrow r<r_h\Rightarrow t=0$, $V=-U>0\Rightarrow sgn(f)>0\Rightarrow r>r_h\Rightarrow t=0.$ $$\eqno{(4.44)}$$ I.e. $t=0$ for both $V=U>0$ ($r<r_h$) and $V=-U>0$ ($r>r_h$). 

\

On the other hand, it is straightforward to verify that $$ds^2_{S\widetilde{AdS}_4}\vert_{t/r}=-{{4f(r)}\over{(f^\prime(r_h))^2}}{{dVdU}\over{UV}}.\eqno{(4.45)}$$

\

From (4.40), $$t=r^*(r)-{{2}\over{f^\prime(r_h)}}ln(-sgn(f)U)=-r^*(r)+{{2}\over{f^\prime(r_h)}}ln(V)$$ and therefore $$r^*(r)={{1}\over{f^\prime(r_h)}}ln(-sgn(f)UV)=r^*(UV)\equiv r^*(U,V)\eqno{(4.46)}$$ with $$r^*(-U,-V)=r^*(U,V).\eqno{(4.47)}$$ Since at each region, $r>r_h$ and $r<r_h$, $r$ can be unambiguously determined from $r^*$, then $r(U,V)=r(-U,-V)$. Then $$(U,V)\to (-U,-V)\eqno{(4.48)}$$ is a symmetry of (4.46) and, moreover, of the entire metric: $$ds^2_{\widetilde{AdS}_4}(U,V)=ds^2_{S\widetilde{AdS}_4}(-U,-V)=-{{4f(r(U,V))}\over{(f^\prime(r_h))^2}}{{dVdU}\over{UV}}-r^2(U,V)d\Omega^2_2.\eqno{(4.49)}$$ This allows the extension of the spacetime to the additional regions $(II)$ ($U=e^{f^\prime(r_h){{r^*(r)-t}\over{2}}}>0$, $V=-e^{f^\prime(r_h){{r^*(r)+t}\over{2}}}<0$) and $(IV)$ ($U=-e^{f^\prime(r_h){{r^*(r)-t}\over{2}}}<0$, $V=-e^{f^\prime(r_h){{r^*(r)+t}\over{2}}}<0$). In $(II)$, ${{U}\over{V}}=-e^{-f^\prime(r_h)t}$; so as $U\to 0_+$ ($-V$-axis), $t\to +\infty$, and as $V\to 0_-$ ($+U$-axis), $t\to -\infty$. In $(IV)$, ${{U}\over{V}}=e^{-f^\prime(r_h)t}$; so as $U\to 0_-$ ($-V$-axis), $t\to +\infty$, and as $V\to 0_-$ ($-U$-axis), $t\to -\infty$. 

\

Finally, to pass from two null coordinates ($U$ and $V$) and two spacelike coordinates ($\theta$ and $\varphi$) to one timelike coordinate and three spacelike coordinates, one defines the {\it Kruskal-Szekeres} coordinates $T$ (timelike), and $X$ (spacelike) through $$U:=T-X, \ V:=T+X. \ \Rightarrow UV=T^2-X^2.\eqno{(4.50)}$$ Let $r>r_h$; for $U<0$ and $V>0$ or $U>0$ and $V<0$, $T^2-X^2=-\vert U\vert V=-U\vert V\vert$ and so $$X=\pm\sqrt{T^2+\vert U\vert V}=\pm\sqrt{T^2+e^{f^\prime(r_h)r^*(r)}}, \ X(T)=X(-T).\eqno{(4.51)}$$ As $r\to {r_h}_+$, $r^*(r)\to -\infty$ and then $$T\to +X \ (future \ horizon), \ T\to -X \ (past \ horizon).\eqno{(4.52)}$$ As $r\to +\infty$, $r^*(r)\to r^*(+\infty)<\infty$ given by (4.39) and then $$X=\pm\sqrt{T^2+e^{f^\prime(r_h)r^*(+\infty)}},\eqno{(4.53)}$$ respectively the {\it right} and {\it left timelike boundaries}. 

\

Let $r<r_h$; for $U,V>0$ or $U,V<0$, $$T=\pm\sqrt{X^2+UV}=\pm\sqrt{X^2+e^{f^\prime(r_h)r^*(r)}}, \ T(X)=T(-X).\eqno{(4.54)}$$ Again, as $r\to {r_h}_-$, $r^*(r)\to -\infty$ and $T=\pm X$. As $r\to 0_+$, $r^*(r)\to 0$ and then $$T\to \pm\sqrt{X^2+1},\eqno{(4.55)}$$ respectively the {\it future} and {\it past spacelike singularities}. 

\

In terms of the coordinate functions $T,X,\theta,\varphi$, the $S\widetilde{AdS}_4$ metric is $$ds^2_{S\widetilde{AdS}_4}(T,X,\theta,\varphi)=-{{4f(r(T,X))}\over{(f^\prime(r_h))^2}}{{dT^2-dX^2}\over{T^2-X^2}}-r^2(T,X)d\Omega^2_2(\theta,\varphi).\eqno{(4.56)}$$ 

\

The complete Kruskal-Szekeres diagram is shown in Fig. 11. Regions $(I)$ and $(II)$ are asymptotically anti-De Sitter, while regions $(III)$ and $(IV)$ are respectively the $SAdS$ black and white holes. In the whole spacetime radial light rays move at $45^o$ and $135^o$, according to ${{dX}\over{dT}}=\pm 1$. 

\begin{figure}[h!]\centering
	\includegraphics[width=\linewidth]{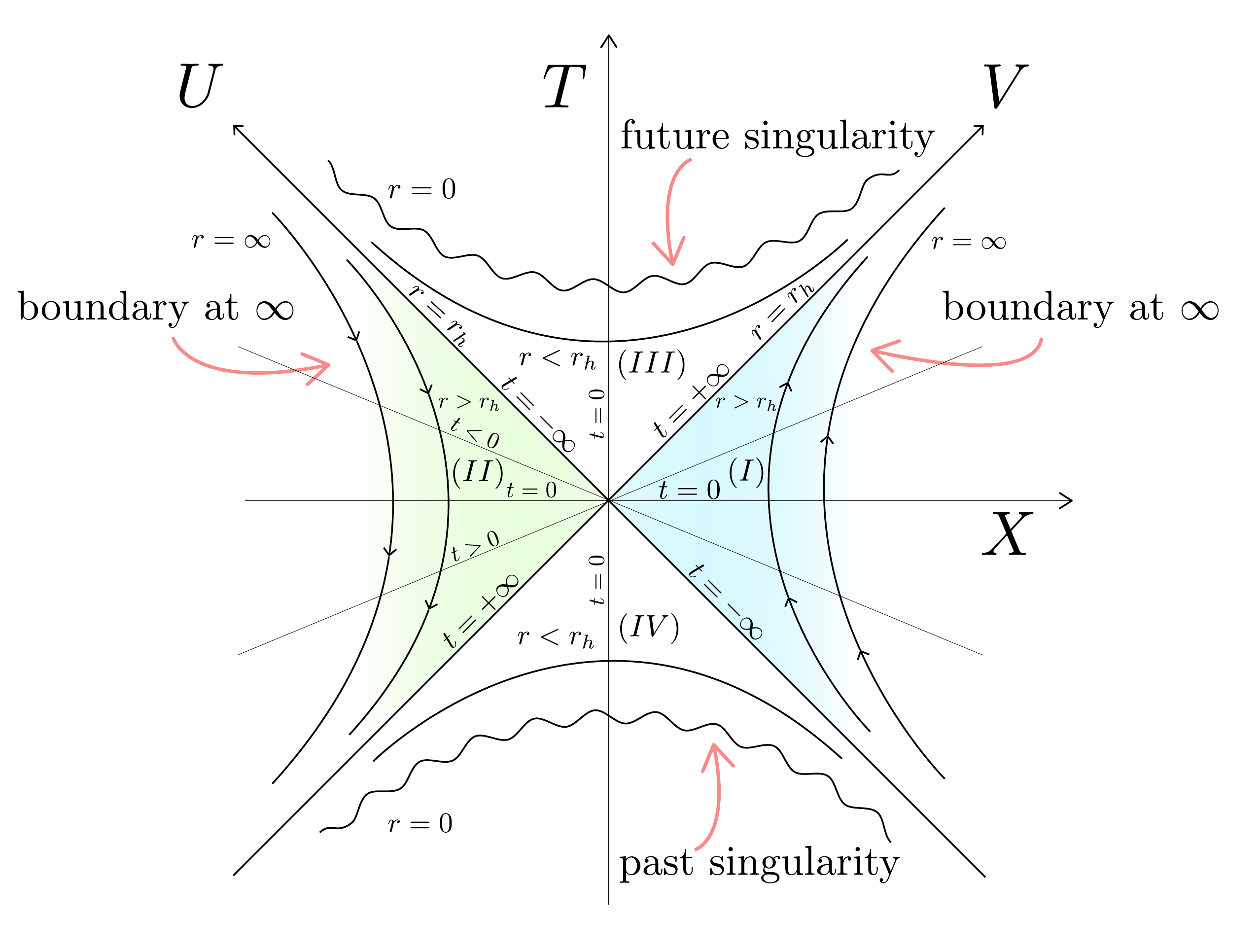}
	\caption{Kruskal-Szekeres diagram for $ \widetilde{SAdS_4} $}
\end{figure}

For $a\to\infty$, the regions $(I)$ and $(II)$ become asymptotically flat and one recovers the pure Schwarzschild results: $$r^*(r)\to r^*_{Schw.}(r)=r+2Mln\vert 1-{{r}\over{2M}}\vert,\eqno{(4.57a)}$$ with $r^*_{Schw.}(+\infty)=+\infty$; $f^\prime(r_h)\to{{1}\over{2M}}$ implying $$T_{Hawk.}\vert_{S\widetilde{AdS}_4}\to T_{Hawk.}\vert_{Schw.}={{1}\over{8\pi M}},\eqno{(4.57b)}$$ and $$X=\pm\sqrt{T^2+({{r}\over{2M}}-1)e^{{{r}\over{2M}}}}\eqno{(4.57c)}$$ for $r>2M$, and  $$T=\pm\sqrt{X^2+(1-{{r}\over{2M}})e^{{{r}\over{2M}}}}\eqno{(4.57d)}$$ for $r<2M$. 

\

A final comment on the implicit presence of the original Schwarzschild coordinates $r$ and $t$ in the metric (4.56): they remain in the form of constant hypersurfaces $r=const.$ and $t=const.$ as can be seen in Fig. 11. In the simpler case of the Schwarzschild black hole ($a=\infty$) this is easily understood through the use of the Eddington-Finkelstein coordinates [17], [18], which allow the extension of $r$ and $t$ to the black hole and white hole regions. The new region (here $II$) is reached only through the introduction of the Kruskal-Szekeres coordinates [19].

\

4.6. {\it Penrose diagram}

\

Define the coordinates $\rho,\tau\in(-\infty,+\infty)$, $[\rho]=[\tau]=L^0$, through $$V:=e^{{{f^\prime(r_h)r^*(+\infty)}\over{2}}}tg({{\rho+\tau}\over{2}}),\eqno{(4.58a)}$$ $$U:=-e^{{{f^\prime(r_h)r^*(+\infty)}\over{2}}}tg({{\rho-\tau}\over{2}}).\eqno{(4.58b)}$$ For $r>r_h$, $f(r)>0$ and from (4.40) $UV=-e^{f^\prime(r_h)r^*(r)}$, at $r=+\infty$ one has $$tg({{\rho+\tau}\over{2}})tg({{\rho-\tau}\over{2}})=1,\eqno{(4.59)}$$ equivalent to $tg({{\rho+\tau}\over{2}})=cotg({{\rho-\tau}\over{2}})$. This implies $cos\rho=0$ which is satisfied by $\rho=\pm\pi/2$. So, $tg({{\pi}\over{4}}+{{\tau}\over{2}})=cotg({{\pi}\over{4}}-{{\tau}\over{2}})$ implies ${{\tau}\over{2}}\in({{-\pi}\over{4}},{{\pi}\over{4}})$ i.e. $\tau\in({{-\pi}\over{2}},{{\pi}\over{2}})$. So, the {\it boundaries} at $r=+\infty$ in the $\tau/\rho$-plane are represented by the timelike straight lines $$\rho=\pm{{\pi}\over{2}}, \  \tau\in({{-\pi}\over{2}},{{\pi}\over{2}}).\eqno{(4.60)}$$

\

From (4.50) and (4.55), for $r=0$, $UV=1=-e^{f^\prime(r_h)r^*(+\infty)}tg({{\rho+\tau}\over{2}})tg({{\rho-\tau}\over{2}})$ which implies $$tg({{\rho+\tau}\over{2}})=l \ cotg({{\rho-\tau}\over{2}}), \ 0<l=e^{-f^\prime(r_h)r^*(+\infty)}=l(M,a)<1.\eqno{(4.61)}$$ For $\rho=0$, $tg({{\tau_0}\over{2}})=l \ cotg({{\tau_0}\over{2}})$ has a unique solution $$0<\tau_0<{{\pi}\over{2}} \ and \ {{-\pi}\over{2}}<\tau^\prime_0=-\tau_0<0.\eqno{(4.62)}$$ For $\rho=+{{\pi}\over{2}}$, $tg({{\tau}\over{2}}+{{\pi}\over{4}})=l \ cotg({{\tau}\over{2}}-{{\pi}\over{4}})$ has the solutions $\tau=\pm{{\pi}\over{2}}$. The same for $\rho=-{{\pi}\over{2}}$. So, by continuity, the {\it future} and {\it past singularities} at $r=0$ are represented by the wavy lines in Fig. 12. 

\begin{figure}[h!]\centering
	\includegraphics[width=\linewidth]{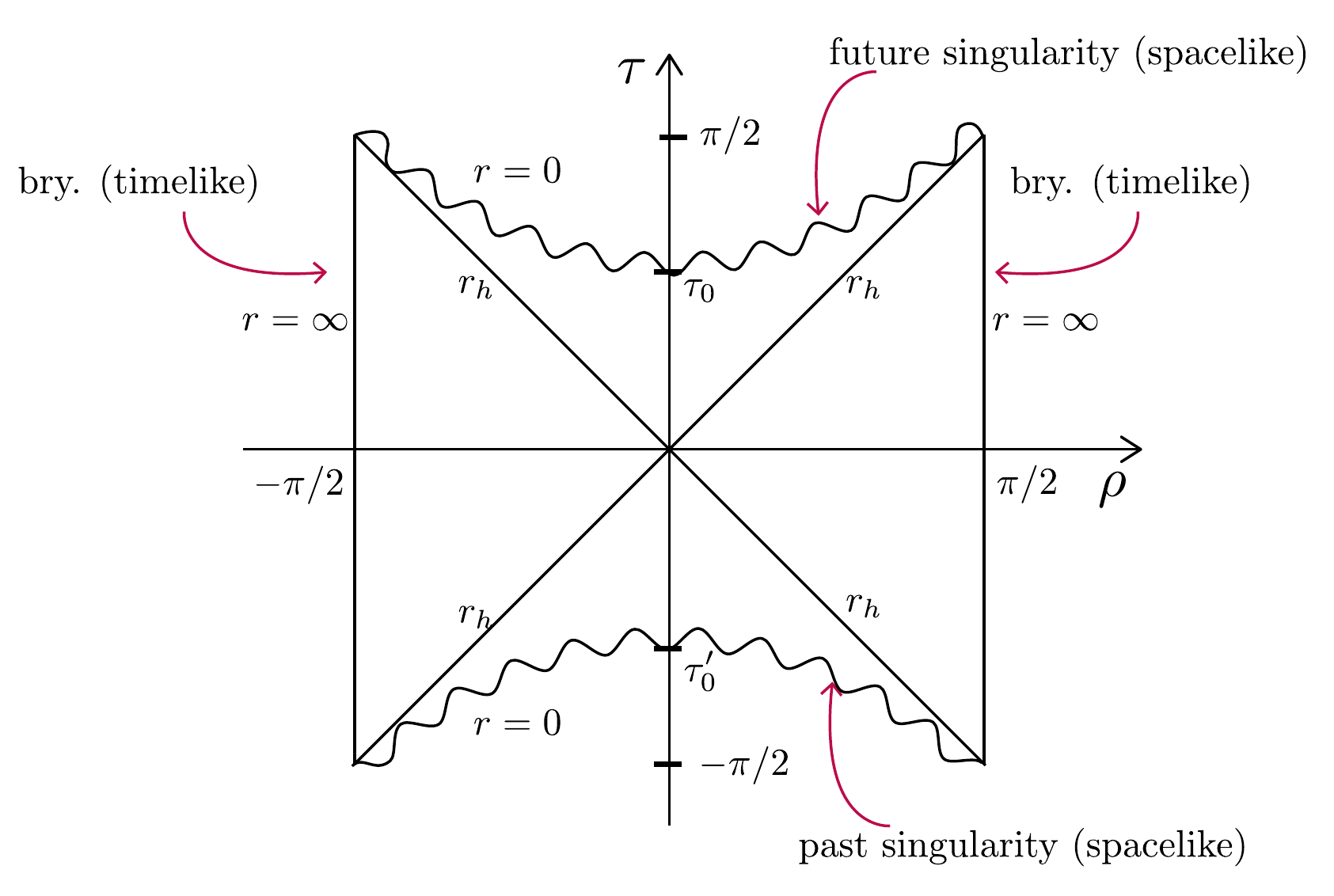}
	\caption{Penrose diagram for $ \widetilde{SAdS_4} $}
\end{figure}

The {\it future} and {\it past horizons} at $r=r_h$ are respectively given by $U=0$ i.e. $tg({{\rho-\tau}\over{2}})=0$ which implies $\rho=\tau$, and by $V=0$ i.e. $tg({{\rho+\tau}\over{2}})=0$ which implies $\tau=-\rho$, the diagonals in the diagram of Fig. 12. 

\

4.7. {\it Entropy}

\

This subsection is a bit more technical, from the physical point of view, than the previous ones. It includes the concepts of partition function and its semiclassical approximation in the language of path integrals; its associated thermodynamic potential, the Helmholtz free energy; and the concept of phase transition. The main objective of the subsection is the computation of the entropy of the $S\widetilde{AdS}_4$ black hole (result (4.84)) following the original derivation of Hawking and Page. Suffice it to say here, is that its counterpart in the Schwarzschild black hole case is much easier to obtain ([3], p. 147), with the same area law behavior. 

\

The partition function (trace of the density matrix in the canonical ensemble) associated with the ${\widetilde{AdS}_4}$ and ${S\widetilde{AdS}_4}$ metrics in thermal equilibrium with a ``heat reservoir" at inverse temperature $\beta={{1}\over{T}}$ is the Euclidean path integral over the whole set of metrics related by coordinate transformations $x^\mu\to {x^\mu}^\prime$, $g_{\mu\nu}(x)\to g^\prime_{\mu\nu}(x^\prime)={{\partial x^\rho}\over{\partial {x^\mu}^\prime}}{{\partial x^\sigma}\over{\partial {x^\nu}^\prime}}g_{\rho\sigma}(x)$: $$Z(\beta)=\int {\cal D}g_{\mu\nu}e^{-I[g_{\mu\nu}]}\eqno{(4.63)}$$ where $$I[g_{\mu\nu}]={{1}\over{16\pi}}\int_0^\beta d\tau \int d^3\vec{x}\sqrt{\vert det(g_{\mu\nu})\vert}{\cal L}_E(g_{\mu\nu})\eqno{(4.64)}$$ is the classical action functional with Lagrangian density $${\cal L}_E(g_{\mu\nu})=R+2\Lambda,\eqno{(4.65)}$$ where here $R$ is the Ricci scalar corresponding to the metric $g_{\mu\nu}$. $d^3\vec{x}=drd\theta d\varphi$, and we have performed a Wick rotation going to imaginary time $t\to -i\tau$ with $\tau\in [0,\beta]$. So $$I[g_{\mu\nu}]={{1}\over{16\pi}}\int_0^\beta d\tau \int d^3\vec{x}\sqrt{\vert det(g_{\mu\nu})\vert}(R+2\Lambda).\eqno{(4.66)}$$ There are two arguments to ``approximate" $Z(\beta)$ by $e^{-I}$:

\

i) In the semiclassical approximation to the theory of black holes, the geometrical objects -in this case the $\widetilde{AdS}_4$ and $S\widetilde{AdS}_4$ spacetimes- are treated classically. Therefore one has to consider the classical limit $\hbar\to 0$ of $Z(\beta)$ where the dominant contribution is the classical action evaluated at the corresponding solution of Einstein equation, the corrections being of $O(\hbar^2)$ (see in this connection refs. [5] and [6]). From (4.8), $$R=-4\Lambda \eqno{(4.67)}$$ and therefore $$I[g_{\mu\nu}]=-{{\Lambda}\over{8\pi}}\int_0^\beta d\tau \int d^3\vec{x}\sqrt{\vert det(g_{\mu\nu})\vert}=-{{\Lambda}\over{8\pi}}\int d {\cal V}ol \eqno{(4.68)}$$ where $d{\cal V}ol$ is the coordinate invariant volume element. One should also consider the Hawking-Gibbons boundary terms [4] for both $\widetilde{AdS}_4$ and $S\widetilde{AdS}_4$; however they cancel each other in the difference between the corresponding classical actions (see below). 

\

ii) Precisely due to (4.68), $e^{-I}$ factors out of the path integral, which becomes the (infinite) constant $\int{\cal D}g_{\mu\nu}$. On taking the logarithm $$ln \ Z(\beta)=-I+ln\int{\cal D}G_{\mu\nu}.\eqno{(4.69)}$$ The second term does not contribute to derivatives of the Helmholtz free energy $$F=-T \ ln \ Z=<E>-TS,\eqno{(4.70)}$$ where $$<E>=-{{\partial}\over{\partial\beta}}ln \ Z\eqno{(4.71)}$$ is the average value of the energy and $$S=\beta <E> +ln \ Z\eqno{(4.72)}$$ is the entropy. The contribution to the entropy of $ln\int{\cal D}G_{\mu\nu}$ is a constant which can be neglected (only entropy differences are important). So we have: $$F=\beta^{-1}I=TI, \eqno{(4.73)}$$ $$<E>={{\partial}\over{\partial\beta}}I, \eqno{(4.74)}$$ and $$S=\beta<E>-I.\eqno{(4.75)}$$ For both $\widetilde{AdS}_4$ and $S\widetilde{AdS}_4$ metrics the invariant volume ${\cal V}ol$ is infinite; we regularize it by taking an infrared cut-off at $r=L$.

\

For $\widetilde{AdS}_4$, from (3.17), $\sqrt{\vert det(g_{\mu\nu}\vert_{\widetilde{AdS}_4})\vert}=r^2sin\theta$ and so $$I_{\widetilde{AdS}_4}=-{{1}\over{8\pi}}\int_0^{\beta_1}d\tau\int_0^Ldrr^2\int d\Omega_2^2=-{{\Lambda}\over{6}}\beta_1L^3,\eqno{(4.76)}$$ with $\beta_1\equiv\beta_{\widetilde{AdS}_4}$.

\

For $S\widetilde{AdS}_4$, from (4.11), with the same angular part as $\widetilde{AdS}_4$ but $r\geq r_h$, $$I_{S\widetilde{AdS}_4}={{\Lambda}\over{6}}\beta(L^3-r_h^3),\eqno{(4.77)}$$ with $\beta\equiv\beta_{S\widetilde{AdS}_4}=T_H^{-1}$. 

\

Since we shall take the limit $L\to +\infty$, the two metrics must coincide at $r=L$ and, in particular, the ``proper time intervals" $\beta_1\sqrt{1+L^2/a^2}$ and $\beta\sqrt{1-2M/L+L^2/a^2}$ must also coincide, i.e. $$\beta_1\sqrt{1+L^2/a^2}=\beta\sqrt{1-2M/L+L^2/a^2}.\eqno{(4.78)}$$ So, for the difference between the two actions one has $$\Delta I=I_{S\widetilde{AdS}_4}-I_{\widetilde{AdS}_4}=-{{\Lambda}\over{6}}\beta(L^3-r_h^3-{{\beta_1}\over{\beta}}L^3)={{\beta}\over{2a^2}}(L^3-r_h^3-L^3\sqrt{{{1-2M/L+L^2/a^2}\over{1+L^2/a^2}}})$$ $$={{\beta}\over{2a^2}}(L^3-r_h^3-L^3\sqrt{1-{{2M/L}\over{1+L^2/a^2}}})={{\beta}\over{2a^2}}(L^3-r_h^3-L^3\sqrt{1-{{2Ma^2}\over{La^2+L^3}}})\simeq {{\beta}\over{2a^2}}(L^3-r_h^3-L^3(1-{{Ma^2}\over{La^2+L^3}}))$$ $$= {{\beta}\over{2a^2}}(-r_h^3+Ma^2)\eqno{(4.79)}$$ as $L\to +\infty$.

\

Using (4.15) one obtains $$\Delta I={{\pi r_h^2}\over {a^2+3r_h^2}}(a^2-r_h^2).\eqno{(4.80)}$$ From (4.34), $$\beta={{4\pi a^2r_h}\over{a^2+3r_h^2}},\eqno{(4.81)}$$then $${{\partial r_h}\over{\partial\beta}}={{1}\over{4\pi a^2}}{{(a^2+3r_h^2)^2}\over{a^2-3r_h^2}}\eqno{(4.82)}$$ and therefore $$<E>_{S\widetilde{AdS}_4}={{\partial}\over{\partial\beta}}\Delta I={{\partial}\over{\partial\beta}}({{\pi r_h^2(a^2-r_h^2)}\over{a^2+3r_h^2}})={{1}\over{2}}r_h(1+{{r_h^2}\over{a^2}})=M\eqno{(4.83)}$$ and $$S_{S\widetilde{AdS}_4}=\beta M-\Delta I=\pi r_h^2={{A}\over{4}}.\eqno{(4.84)}$$ As for the pure Schwarzschild case, one obtains for the entropy of the Schwarzschild anti-De Sitter black hole one fourth of the horizon area A. 

\

From (4.80) we notice that $$T\Delta I=T(I_{S\widetilde{AdS}_4}-I_{\widetilde{AdS}_4})=F_{S\widetilde{AdS}_4}-F_{\widetilde{AdS}_4}={{\pi r_h^2T}\over{a^2+3r_h^2}}(a^2-r_h^2)\eqno{(4.85)}$$ vanishes for $r_h=a$ i.e. when $$T=T_1={{1}\over{\pi a}}.\eqno{(4.86)}$$ From (4.35) and (4.86) $$T_0={{\sqrt{3}}\over{2}}T_1<T_1.\eqno{(4.87)}$$ Then, for $T_0<T<T_1$ or, equivalently, for ${{a}\over{\sqrt{3}}}<r_h<a$, $\Delta I>0$ i.e. $$F_{S\widetilde{AdS}_4}>F_{\widetilde{AdS}_4}\eqno{(4.88)}$$ and therefore, though $S\widetilde{AdS}_4$ black holes coexist with pure $\widetilde{AdS}_4$, the second spacetime is preferable, while for $T_1<T$ i.e. $a<r_h$, $\Delta I<0$ i.e. $$F_{S\widetilde{AdS}_4}<F_{\widetilde{AdS}_4},\eqno{(4.89)}$$ and black holes dominate over anti-De Sitter space. $T_1$ is then the temperature at which it occurs the Hawking-Page phase transition [4].

\

\section{Final comments}

\

The Schwarzschild anti-De Sitter metric, which includes both a black hole and a white hole, and two asymptotically anti-De Sitter causally unconnected spacetimes, is an extraordinary laboratory to study, at least theoretically, black hole physics. In particular, black hole thermodynamics, which we roughly reviewed in this article, leads, even in a semiclassical (non quantum) treatment, to a non vanishing entropy proportional to the area of the event horizon (eq. (4.84)). The microscopic quantum description begun with Maldacena's thesis [20] in the context of string theory [21] and led to important developments like the $AdS/CFT$ conjecture [2]. A good review of the thermodynamic aspects can be found in ref. [22]. The ${S\widetilde{AdS}_4}$ black hole is also briefly discussed in Susskind-Lindesay [23], in the context of information theory, holography, and string theory. 

\

{\bf Acknowledgements}

\

The author thanks O. Brauer for drawing the figures, and H. A. Camargo and E. Eiroa for useful discussions. Also to IAFE-UBA-CONICET for its hospitality.

\

{\bf References}

\

[1] Bengtsson, I. {\it Anti-De Sitter Space}, Lecture Notes (1998).

\

[2] Maldacena, J. The Large-N Limit of Superconformal Field Theories and Supergravity, Int. Jour. Theor. Phys. {\bf 38}, 1113-1133 (1998).

\

[3] Carroll, S. {\it Spacetime and Geometry. An Introduction to General Relativity} (Addison-Wesley, San Francisco, 2004), pp. 139-144.

\

[4] Hawking, S.W. and Page, D.N. Thermodynamics of Black Holes in Anti-de Sitter Space, Comm. Math. Phys. {\bf 87}, 577-588 (1983).

\

[5] Zhao, P. {\it Black Holes in Anti-de Sitter Spacetime}, Lent term Part III Seminar Series, Essay (2008).

\

[6] Charmousis, C. Introduction to Anti de Sitter Black Holes, Chapter 1, in {\it From Gravity to Thermal Gauge Theories: the AdS/CFT Correspondence}, Lecture Notes in Physics {\bf 828}, (2011).

\

[7] Kristiansson, F. An Excursion into the Anti-de Sitter Spacetime and the World of Holography, Master Degree Project, Uppsala University (1999).

\

[8] Maldacena, J. The Illusion of Gravity, Scient. Am. 75-81, april 2007.

\

[9] 't Hooft, G. Introduction to the Theory of Black Holes, ITP-SPIN (2009), pp. 31-32.

\

[10] Socolovsky, M. Rindler space, Unruh effect and Hawking temperature, Annales de la Fondation Louis de Broglie {\bf 39}, 1-49 (2014).

\

[11] Unruh, W.G. Notes on black hole evaporation, Phys. Rev. D {\bf 14}, 870-892 (1976).

\

[12] Hawking, S.W. Particle Creation by Black Holes, Comm. Math. Phys. {\bf 43}, 199-220 (1975).

\

[13] Wald, R. {\it General Relativity}, The University Chicago Press, Chicago and London (1984), p. 152.

\

[14] Kruskal, M.D. Maximal extension of Schwarzschild Metric, Phys. Rev. {\bf 119}, 1743-1745 (1960).

\

[15] Szekeres, G. On the singularities of a Riemannian manifold, Publ. Math. Debrecen {\bf 7}, 285-301 (1960).

\

[16] Kloesch, T. and Strobl, T. Classical and Quantum Gravity in 1+1 Dimensions, Part II: The Universal Coverings, Class. Quant. Grav. {\bf 13}, 2395-2422 (1996); arXiv: gr-qc/9511081v3.

\ 

[17] Eddington, A.S. A comparison of Whitehead's and Einstein's Formulae, Nature {\bf 113}, 192 (1924).

\

[18] Finkelstein, D. Past-Future Asymmetry of the Gravitational Field of a Point Particle, Phys. Rev. {\bf 110}, 965-967 (1958).

\

[19] Hawking, S.W. and Ellis, G.F.R. {\it The Large Scale Structure of Space-Time}, Cambridge University Press, Cambridge (1973), pp. 146-154.

\

[20] Maldacena, J.M. {\it Black Holes in String Theory}, PhD Thesis, Princeton University (1996).

\

[21] Zwiebach, B. {\it A First Course in String Theory}, Cambridge University Press, Cambridge, 2nd. ed. (2009).

\

[22] Jacobson, T. {\it Introductory Lectures on Black Hole Thermodynamics}, Inst. Theor. Phys., University Utrecht.

\

[23] Susskind, L. and Lindesay, J. {\it An Introduction to Black Holes, Information, and the String Revolution. The Holographic Universe}, World Scientific, Singapore (2005).

\

{\bf Contents}

\

1. {\it Introduction}

\

2. {\it The hyperbolic plane}

\

2.1. Pseudoeuclidean space $\mathbb{E}^{(2,1)}$; metric in coordinates $X,Y,Z$

\

2.2. Hyperbolic plane ($H.P.$): half of two-sheets hyperboloid with curvature radius $a$

\

2.3., 2.4. Global coordinates $\rho\in [0,+\infty)$, $\varphi\in [0,2\pi)$

\

2.5. Embedding of $H.P.$ in $\mathbb{E}^{(2,1)}$

\

2.6. Metric of $H.P.$

\

2.7. Poincar\'e projection (Poincar\'e disk), metric. Conformal equivalence $H.P.\vert_{Poinc.}\buildrel \ {conf.}\over\cong H.P.$

\

2.8. Metric of $H.P.$ in coordinates $\chi\in [0,\pi/2)$, $\varphi$

\

2.9. Metric of $H.P.$ in coordinates $r\in [0,+\infty)$, $\varphi$

\

2.10. Metric of $H.P.$ in Poincar\'e coordinates $\tau$, $\iota$ and $\tau$, $s$

\

2.11. Poincar\'e half plane: coordinates $x\in (-\infty,+\infty)$, $y>0$

\

2.12. Scalar curvature of $H.P.$

\

2.13. Vertical distance between points

\

2.14. Horizontal distance between points

\

2.15. Three dimensional hyperbolic space $H^3$

\

\

3. {\it Four dimensional anti-De Sitter spacetime ($AdS_4$) and its universal covering $\widetilde{AdS}_4$}

\

3.1. Pseudoeuclidean space $\mathbb{E}^{(2,3)}$

\

3.2. $AdS_4\subset\mathbb{E}^{(2,3)}$

\

3.3., 3.4. Metric of $AdS_4$ in global coordinates $t^\prime\in[0,2\pi]$, $\rho\in[0,+\infty)$, $\theta\in[0,\pi]$, $\varphi\in[0,2\pi)$

\

3.5. Symmetry groups of $\widetilde{AdS}_4$, $SO(2,3)=Conf(Mink^3)$, and of $\widetilde{AdS}_5$, $SO(2,4)=Conf(Mink^4)$

\

3.6. Conformal Penrose diagram

\

3.7. Boundary of $AdS_d$ and of $\widetilde{AdS}_d$

\

3.8. Spherical or static coordinates

\

3.9. Scalar curvature of $\widetilde{AdS}_4$ and cosmological constant $\Lambda=-{{3}\over{a^2}}$

\

3.10. Coordinates for $\widetilde{AdS}_4/2$

\

3.11. Poincar\'e coordinates. Scale invariance. Example of $\widetilde{AdS}_3$: massless and massive radial geodesics

\

3.12. $\widetilde{AdS}_4$ as a stack of $H^3$'s

\

3.13. Tortoise radial coordinate $r^*(r)$

\

\

4. {\it Four dimensional Schwarzschild anti-De Sitter ($S\widetilde{AdS}_4$) metric}

\

4.1. Schwarzschild coordinates

\

4.2. Horizon $r_h(M,a)$ (explicit calculation)

\

4.3. Surface gravity $\kappa$

\

4.4. Rindler approximation at horizon and Hawking temperature $T_{Hawk.}={{\kappa}\over{2\pi}}={{1}\over{2\pi}}({{M}\over{r_h^2}}+{{r_h}\over{a^2}})$

\

4.5. Tortoise radial coordinate $r^*(r)$ (explicit calculation). Kruskal-Szekeres coordinates: $(U,V)$, $(T,X)$. Complete $K-S$ diagram: asymptotically $\widetilde{AdS}_4$, black and white hole regions. Singularity, horizon and boundary.

\

4.6. Penrose diagram

\

4.7. Entropy

\

5. {\it Final comments}

\

Acknowledgments

\

References

\

\

\

\

\

\

\

\

\

$^*$ {\it With a leave of absence from Instituto de Ciencias Nucleares, Universidad Nacional Aut\'onoma de M\'exico}

\

e-mails: msocolov@ungs.edu.ar, socolovs@nucleares.unam.mx

\

\end{document}